\documentclass[%
twocolumn,
amsmath,amssymb,
aps,pra,
superscriptaddress,
floatfix
]{revtex4-2}
\usepackage{amsmath}
\usepackage{graphicx}
\usepackage{bm}
\usepackage{hyperref}
\hypersetup{
colorlinks=true,
filecolor=true,
urlcolor=blue,
citecolor=blue,
linkcolor=blue,
}
\usepackage{booktabs}

\begin{document}

\title{Topological phases and edge modes of an uneven ladder}
\author{Wen-Chuang Shang}
\affiliation{%
Department of Physics, Zhejiang Normal University, Jinhua, 321004, China
}
\affiliation{Key Laboratory of Optical Information Detection and Display Technology of Zhejiang, Zhejiang Normal University, Jinhua, 321004, China}
\author{Yi-Ning Han}
\affiliation{%
Department of Physics, Zhejiang Normal University, Jinhua, 321004, China
}
\affiliation{Key Laboratory of Optical Information Detection and Display Technology of Zhejiang, Zhejiang Normal University, Jinhua, 321004, China}
\author{Shimpei Endo}
\affiliation
{Department of Engineering Science, The University of Electro-Communications, Tokyo 182-8585, Japan}
\author{Chao Gao}
\email{gaochao@zjnu.edu.cn}
\affiliation{Department of Physics, Zhejiang Normal University, Jinhua, 321004, China
}
\affiliation{Key Laboratory of Optical Information Detection and Display Technology of Zhejiang, Zhejiang Normal University, Jinhua, 321004, China}

\date{\today}

\begin{abstract}
We investigate the topological properties of a two-chain quantum ladder with uneven legs, i.e. the two chains differ in their periods by a factor of two. 
Such an uneven ladder presents rich band structures classified by the closure of either  direct or indirect bandgaps. 
It also provides opportunities to explore fundamental concepts concerning band topology and edge modes, including the difference of intracellular and intercellular Zak phases, and the role of the inversion symmetry (IS). 
We calculate the Zak phases of the two kinds and find excellent agreement with the dipole moment and extra charge accumulation, respectively. 
We also find that configurations with IS feature a pair of degenerate two-side edge modes emerging as the closure of the direct bandgap, while configurations without IS feature one-side edge modes emerging as not only the closure of both direct and indirect bandgap but also within the band continuum. 
Furthermore, by projecting to the two sublattices, we find that the effective Bloch Hamiltonian corresponds to that of a generalized Su–Schrieffer–Heeger model or Rice-Mele model whose hopping amplitudes depend on the quasimomentum. 
In this way, the topological phases can be efficiently extracted through winding numbers. 
We propose that uneven ladders can be realized by spin-dependent optical lattices and their rich topological characteristics can be examined by near future experiments.

\


\

\end{abstract}

\maketitle

\section{Introduction}
Topological order has been an active theme in various branches of physics over the past decades. 
Among different topological matters, a unified paradigm focuses on the symmetry-protected topological~(SPT) phases \cite{Chen_2012}. Distinct SPT phases cannot be smoothly transformed into each other without closing the gap if the Hamiltonian respects certain protected global symmetries.
A nontrivial SPT phase is different from a trivial one due to the existence of nontrivial edge states on open boundaries, and its  bulk topological invariant is related to the number of the topological edge states, dubbed bulk-boundary correspondence~(BBC)~\cite{Essin2011,Mong2011,Rhim2017,Rhim2018}.

A prominent demonstration of SPT matters in one dimension is the  Su–Schrieffer–Heeger~(SSH) model~\cite{Su_1979} which is simply a single bipartite chain with two alternating hopping amplitudes. It hosts two different SPT phases distinguished by the relative strength of intracellular and intercellular hopping and such phases are protected by chiral symmetry and (spatial) inversion symmetry (IS).
Bulk topological properties therein can be described by invariants such as winding number~\cite{Asb_th_2016} or Zak phase~\cite{zak1989berry}. 
There exists a pair of degenerate zero-energy two-side edge states in the nontrivial SPT phase corresponding to nonzero topological invariants.   
By further including an on-site staggered potential to the SSH model, one obtains the Rice-Mele~(RM) model~\cite{Rice_Mele_1982}, which breaks both chiral symmetry and IS. 
The winding number is not well-defined therein, and the Zak phase for each energy band is not quantized, implying the absence of SPT phases. 
However, a pair of edge states, although being non-degenerate and chiral, still exist in the same parameter regimes same as that of the SSH model~\cite{ManXin2019,Lin_2020}.

Since the discovery of topological matters, intensive efforts have been devoted to more complicated architectures, among which, topological ladders, consisting of several coupled chains, become a focus. Due to the quasi-one-dimensional (1D) nature and compatibility with various experimental tools, 
versatile topological properties have been explored in ladder architectures by further including 
interactions~\cite{19Science,20PRB,24PRB}, 
orbital degree~\cite{13PRA,13NC,18PRL}, 
spin degree~\cite{03PRB,12PRB,13PRB,19PRL_spin}, 
disorder~\cite{18PRX,22PRL_disorder}, 
or implementing spin-orbit coupling and synthetic gauge field~\cite{16PRL,22LSA}, 
or engineering flat bands~\cite{17PRX,21PRL_flat_a,21PRL_flat_b}. 
Recently, nonequilibrium dynamics~\cite{14NP,19PRL_chiral,22NC,23NC}  and quantum Hall signatures~\cite{16PRA,19PRL_Hall_a,19PRL_Hall_b,21PRL_Hall,22PRL_Hall,22PRXQuantum,22PRXQuantum,23Science} induced by topology have also been investigated in ladder systems both theoretically and experimentally.

Provious studies on the ladder architectures commonly set the legs as even, i.e. the composing chains possess the same periodicity, at most differ by certain parallel translation~\cite{17PRA,18PRX,19PRL_chiral,20PRB,23NC}. 
Releasing the constraint on the identity of the composing chains will potentially enable new perspectives on topological characteristics.
Motivated by such a scenario, in this paper, we propose a new type of ladder model with uneven legs. 
For demonstration, we investigate a minimal uneven ladder model with two legs whose periods differ by a factor of two.
Such an uneven ladder provides opportunities to explore fundamental topological properties. 
Our main results can be outlined as follows. 

(i) We map this two-chain ladder to a  single tripartite chain with hopping terms up to fourth order in the tight-binding limit. 
Under periodic boundary conditions, this model features three basic bands, with either direct or indirect gaps in between.
We classify its band structure according to the closure and reopening of these gaps. 

(ii) By projecting to the two sublattices, we find that the effective Bloch Hamiltonian corresponds to that of a generalized SSH model or RM model whose hopping amplitudes depend on the quasimomentum. In this way, the topological phases can be efficiently extracted through winding numbers. 

(iii) We calculate both the intracellular and intercellular Zak phases~\cite{Rhim2017}. These two phases are then found to be in excellent agreement with the dipole moment and extra charge accumulation, respectively. 

(iv) Under open boundary conditions, we find different types of edge states that are closely related to the existence of the IS.  Configurations with IS feature a pair of degenerate two-side edge modes emerging as the closure of the direct bandgap, while configurations without IS feature chiral edge modes emerging as not only the closure of both direct and indirect bandgap but also within the band continuum. 

The paper is organized as follows. 
Sec.~\ref{model} introduces the model Hamiltonian and discusses its symmetries,  as well as band structure classified by the closure of bandgaps. 
Sec.~\ref{projection} presents the projection of the ladder model to the generalized SSH model. 
Sec.~\ref{Zak} discusses the two kinds of Zak phases and their relation to the dipole moment and  extra charge accumulation. 
Sec.~\ref{edge} presents the edge states and discusses BBC. 
Sec.~\ref{extension} outlooks on more generic two-leg uneven ladders. 
Finally Sec.~\ref{conclusion} presents conclusions.

\section{uneven ladder model} \label{model}

Uneven ladders are composed of several chains with different periods and hopping amplitudes, 
and their periods can be either commensurate or incommensurate.
For simplicity, here we consider a two-leg ladder, and the composing chains are just simple ones with only the nearest-neighbor hopping.  
Furthermore, we choose the periods of the two chains differing by a factor of two.
We note that more complicated configurations can be constructed and the physical properties can be discussed based on analysis in this paper. 

\begin{figure}[t]
\includegraphics[width=8.6cm]{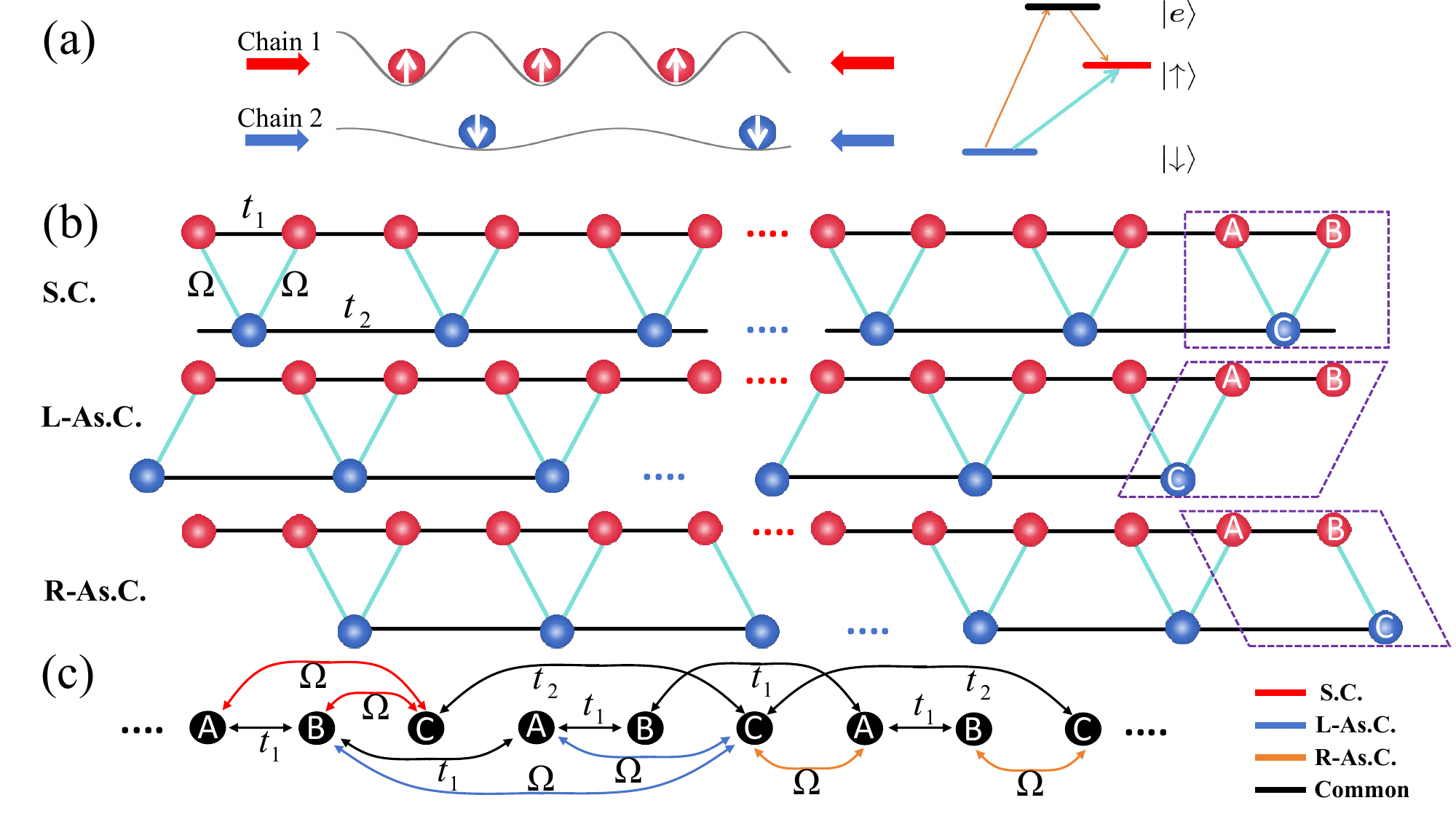}
\caption{
 (a) Experimental scheme of a two-leg uneven ladder constructed by spin-dependent optical lattices and microwaves. 
 (b) Schematic of three ladder configurations: symmetric configuration~(S.C.), left-asymmetric configuration~(L-As.C.), and right-asymmetric configuration~(R-As.C.). 
 These configurations are the same in the bulk but are different at the edges and they correspond to three different kinds of unit cells as shown by purple dashed boxes. 
 (c) Illustration of the mapped single chain equivalent to the uneven ladder. 
 Different colored lines represent couplings of different configurations. 
 }\label{setupfig}
\end{figure}

The ladder architectures can be realized in experimental platforms including photonics~\cite{22NC}, superconductor circuits~\cite{21PRL_flat_b}, thermal atoms~\cite{19PRL_chiral,21PRL_flat_a,22PRL_disorder}, and cold atoms~\cite{14NP,16PRL,17SciAdv,18PRL,18PRX,19Science,22LSA}, etc. 
Uneven ladders can be implemented by tools with separate control of each leg. 
In terms of cold atoms, ladders can be possibly realized by spin-dependent optical lattices~\cite{07PRA,11PRA,wen2021}. 
Such a technology has already been proven to be successful in realizing twisted-bilayer lattices~\cite{Meng_2023}. 
To implement our minimal ladder model in cold atoms, one can  first  impose a pair of counterpropagating laser beams, which confine atoms of two internal spins (denoted as $\uparrow$ and $\downarrow$) independently, such that the two leg chains~(denoted as $1$ and $2$ respectively) are created. 
The periods of lattices for each spin are taken as $a/2$ and $a$, respectively. 
The other two spatial degrees can be frozen out by extra strong optical lattices or  optical dipole traps, such that the system is quasi-1D. 
The two spins can be further coupled by microwave~(or radio-frequency wave depending on the spin-flip frequency) such that the two chains are effectively linked.
The proposed experimental scheme and the corresponding ladder configurations are illustrated in Fig.~\ref{setupfig}(a, b).
 
In the tight-binding limit, the Hamiltonian of the minimal two-leg uneven ladder reads
\begin{eqnarray}
H=&-t_1&\sum_{j_1}^{} (a_{j_1}^\dagger a_{j_1+1}+ \text{H.c.})-t_2\sum_{j_2}^{} (b_{j_2}^\dagger b_{j_2+1}+ \text{H.c.}) \nonumber \\ &+&\Omega\sum_{{\left<j_1,j_2\right>}}^{}(a_{j_1}^\dagger b_{j_2}+\text{H.c.}),
\end{eqnarray}
where $a_{j_1}(a_{j_1}^{\dagger})$ and $b_{j_2}(b_{j_2}^{\dagger})$ are creation~(annihilation) operators at site $j_1$ and $j_2$ of chain-$1$ and chain-$2$ respectively,  $t_1$ and $t_2$ are the intra-chain hopping amplitudes, and $\Omega$ is the inter-chain coupling strength.
Due to the exponential decay of hopping/coupling strengths as the increase of site distance, here we can consider only the nearest-neighbor site  hopping and coupling. 
The uneven ladders can be mapped to a single chain. 
Here, our minimal ladder model is equivalent to a tripartite single chain with the subsite denoted as $A$, $B$, $C$.  The hopping amplitudes of such a chain count up to the fourth order, as shown in Fig.~\ref{setupfig}(c). 
The mapped chain belongs to the general class of trimer SSH models while goes beyond previous studies~\cite{Su_1981,Guo_2015, Alvarez2019,22PRB} due to the inclusion of long-range hoppings. 

\subsection{Ladder configurations and symmetries\label{symmetry}}

In the SSH model, different choices of the origin of unit cells lead to different kinds of topological phases and different behavior of edge states, even though the system parameters are taken the same~\cite{13NP}. 
Here, we demonstrate that the two-leg uneven ladder enables more configurations of unit cells, thus supporting more peculiar topological properties. 
The ladder configurations basically can be classified into three types as shown in Fig.~\ref{setupfig}(b): 
symmetric configuration~(S.C.), left-asymmetric configuration~(L-As.C.), and right-asymmetric configuration~(R-As.C.). 
These configurations are the same in the bulk but are different at the edges and the unit cells. 

Naturally, one can adopt a unit cell with three subsites including adjacent $A$ and $B$  in chain-$1$ and $C$ site in chian-$2$ just below $A$ and $B$. This is the so-called S.C., see the first row in Fig.~\ref{setupfig}(b). 
The corresponding Hamiltonian reads
\begin{eqnarray}
\begin{aligned}
H_{\text{SC}} =&-t_1\sum_{m}(a_{A,m}^\dagger a_{B,m}+a_{A,m+1}^\dagger a_{B,m}+\text{H.c.}) \\
&-t_2\sum_{m}(a_{C,m+1}^\dagger a_{C,m}+\text{H.c.}) \\
&+\Omega \sum_{m}(a_{C,m}^\dagger a_{A,m}+a_{C,m}^\dagger a_{B,m}+\text{H.c.}), \label{scobcH}    
\end{aligned}
\end{eqnarray}
where $a_{X,m}^\dagger$ ($a_{X,m}$) is the creation (annihilation) operator at the subsite $X$ ($X$ can be $A$, $B$, $C$) of the $m$-th unit cell.
Under periodic boundary conditions (PBC), by performing a Fourier transformation $a_{X,k}=1/\sqrt{M_t} \sum_m e^{ikm}a_{X,m}$, where $M_t$ is the total number of unit cells in the system, one obtains the Bloch Hamiltonian:
\begin{equation}
H_{\text{SC}}(k) =\begin{pmatrix}
0 & -t_1-t_1e^{-ik} & \Omega \\
-t_1-t_1e^{ik} & 0 & \Omega \\
\Omega & \Omega & -2t_2 \cos k
\end{pmatrix}.
\end{equation}
By adopting another unit cell, with $C$ site in chain-$2$ left to the site $A$, one obtains the so-called L-As.C.. 
The corresponding Hamiltonian in real space and the Bloch Hamiltonian read
\begin{eqnarray}
\begin{aligned}
H_{\text{LA}}=&-t_1\sum_{m}(a_{A,m}^\dagger a_{B,m}+a_{A,m+1}^\dagger a_{B,m}+\text{H.c.}) \\
&-t_2\sum_{m}(a_{C,m+1}^\dagger a_{C,m}+\text{H.c.})  \\
&+\Omega \sum_{m}(a_{C,m}^\dagger a_{A,m}+a_{C,m+1}^\dagger a_{B,m}+\text{H.c.}) ,    
\end{aligned}
\end{eqnarray}
\begin{equation}
H_{\text{LA}}(k) =\begin{pmatrix}
0 & -t_1-t_1e^{-ik} & \Omega \\
-t_1-t_1e^{ik} & 0 & \Omega e^{ik}\\
\Omega & \Omega e^{-ik} & -2t_2 \cos k
\end{pmatrix},
\end{equation}
respectively. Similarly, the Bloch Hamiltonian of R-As.C. reads
\begin{equation}
H_{\text{RA}}(k) =\begin{pmatrix}
0 & -t_1-t_1e^{-ik} & \Omega e^{-ik} \\
-t_1-t_1e^{ik} & 0 & \Omega \\
\Omega e^{ik}& \Omega & -2t_2 \cos k
\end{pmatrix}.
\end{equation}
It should be noted that, under PBC, Bloch Hamiltonians of different configurations are equivalent to each other up to unitary transformation. 
One can find that $U H_{\text{SC}}(k) U^{\dagger}=H_{\text{LA}}(-k)$ and ${U^{\prime}} H_{\text{SC}}(k){U^{\prime}}^{\dagger}=H_{\text{RA}}(-k)$, where
\begin{eqnarray*}
U=\begin{pmatrix}
1 & 0& 0 \\
0 & e^{-ik} & 0\\
0 & 0 & 1
\end{pmatrix},
U^{\prime}=\begin{pmatrix}
e^{ik}& 0& 0 \\
0 & 1 & 0\\
0 & 0 & 1
\end{pmatrix}, \label{7000}
\end{eqnarray*}
are unitary matrices,  $UU^\dagger=U^{\prime}{U^{\prime}}^\dagger=I$. 
However, under open boundary conditions (OBC), these configurations are not equivalent. 
It is obvious to view their difference at the edges as shown in Fig.~\ref{setupfig}(b). 

In what follows, we discuss the symmetries of the two-leg uneven ladder of different configurations, mainly time-reversal symmetry, chiral symmetry and IS.

Firstly, we can readily verify that the uneven ladder model possesses time-reversal symmetry, i.e., $\tau H(k) \tau^{-1} = H(-k)$, where $\tau$ represents the time-reversal operator. 
For spinless systems, the time-reversal operator results in the overall complex conjugation of the Hamiltonian \cite{Griffiths_2018}.

\begin{figure*}[t]
\includegraphics[width=17.9cm]{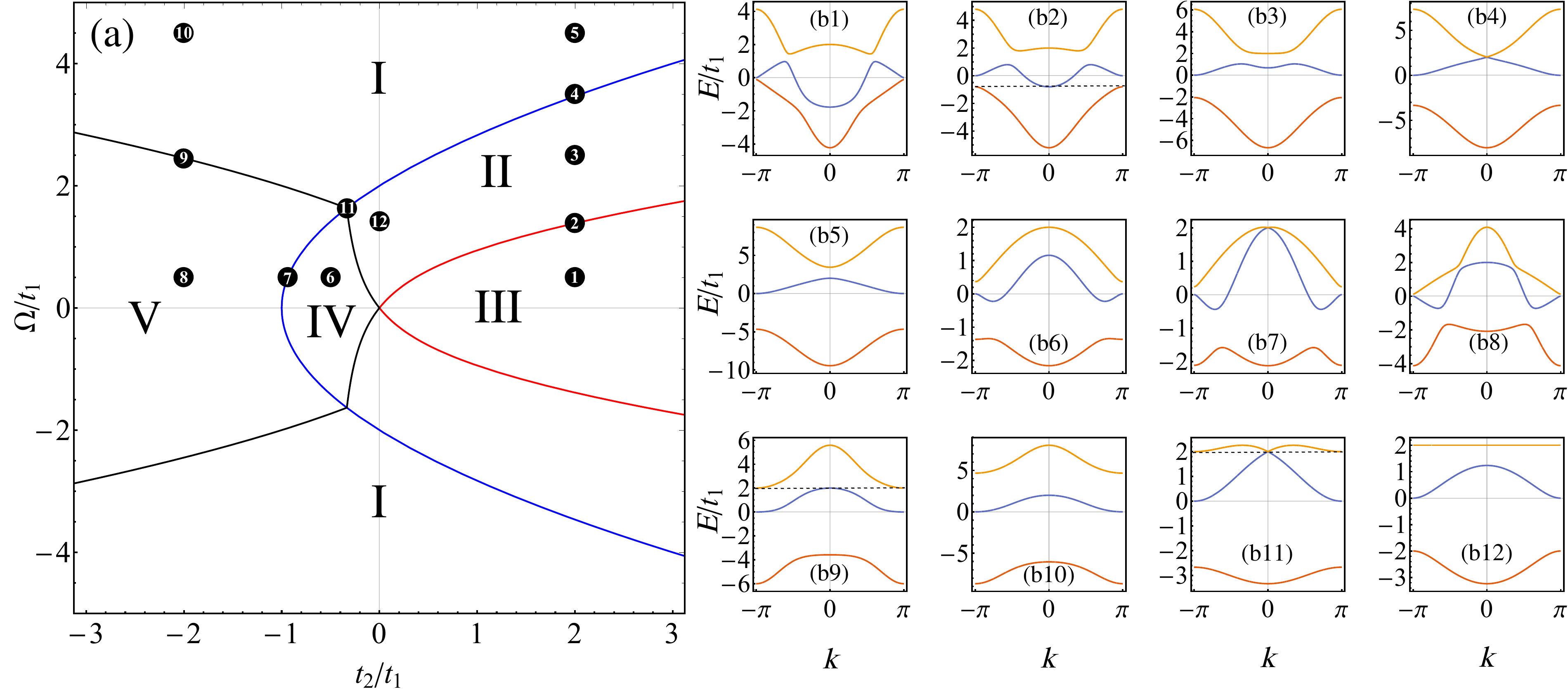}
\caption{
(a) Bandgap closure scenarios. 
Lines denote gap closure between different bands, and divide the parameter space into five regions accordingly. 
For $t_1 > 0$, the blue line indicates the direct bandgap closure between the second and third bands, the red line indicates the indirect bandgap closure between the first and second bands, and the black line indicates the indirect bandgap closure between the second and third bands. 
(b1-b12) typical band structures corresponding to the labels in (a).
The dashed black lines denote the position where indirect band closures. 
When $t_1 < 0$, the band structure undergoes a global inversion relative to the case for $t_1 > 0$ (symmetric about $E=0$), implying the exchange of  the first and third bands. 
 for (b1-b5), the parameters are set as $t_2/t_1 = 2$, and  $\Omega/t_1 = 4.5$, $3.5$, $2.5$, $4 \sqrt{3}/5$, $0.5$, respectively. 
For (b6-b8),  $\Omega/t_1 = 0.5$, and $t_2/t_1 = -0.5$, $-0.938$, $-2$, respectively. 
For (b9-b10),  $t_2/t_1 = -2$,  $\Omega/t_1 = \sqrt{6}$, $4.5$, respectively. 
For (b11), corresponding to the intersection of the blue and black lines in (a),  $t_2/t_1 = -1/3$ and $\Omega/t_1 = 2 \sqrt{2/3}$. 
For (b12),  where a flat band with energy $2t_1$ emerges,  $t_2 = 0$ and $\Omega/t_1 = \sqrt{2}$.} \label{band structure}
\end{figure*}

It is obvious that the chiral symmetry is absent for the uneven ladder, as can be shown in Fig.~\ref{band structure} for PBC and \ref{t1t2morethan0}(a) for OBC that the energy spectrum is not symmetric against any reference energy. 
It means that no operator, denoted as $C$, can be found to ensure $CHC^\dagger=-H$. 

It is worth noting that, in a previous study on a trimer SSH model with only nearest-neighbor hopping~\cite{22PRB}, a generalized chiral symmetry, called point chirality, was discovered.  
The corresponding Bloch Hamiltonian there satisfies $\Gamma H(k)\Gamma^\dagger=-H(2k_0+k)$, with $\Gamma$ being the matrix for point chirality. 
The energy spectrum then is symmetric against a reference energy for OBC, 
and also symmetric in the momentum-energy~($k,E$) diagram for PBC with respect to a certain point,  ($k=k_0,E=0$).
However, in the uneven ladder model discussed here, which is equivalent to a trimer SSH chain, such point chirality is broken down by the long-range hoppings.

Concerning the IS, the scenario for the uneven ladder is different from the chiral symmetry. 
Under PBC, the uneven ladder of any configuration exhibits IS. 
Specifically, we can ensure that the Bloch Hamiltonians of different configurations all satisfy the IS, $P_0 H_{\text{SC}}^{\text{}}(k)P_0^\dagger=H_{\text{SC}}(-k)$, $P_{+}H_{\text{LA}}^{\text{}}(k)P_{+}^\dagger=H_{\text{LA}}(-k)$, and $P_{-} H_{\text{RA}}(k)P_{-}^\dagger=H_{\text{RA}}(-k)$. 
Here, the operators $P_0$  and $P_{\pm}$, standing for the IS, read
\begin{eqnarray*}
P_0=\begin{pmatrix}
0 & 1& 0 \\
1& 0 & 0\\
0 & 0 & 1
\end{pmatrix},
P{\pm}=\begin{pmatrix}
0 & 1& 0 \\
1& 0 & 0\\
0 & 0 & e^{\pm ik}
\end{pmatrix},
\end{eqnarray*}
and can be understood as swapping sublattices $A$ and $B$ while maintaining sublattice $C$ or just adding some phases. 
Notice that $P_0$ is independent of the momentum $k$, while $P_\pm$ depends on $k$. 
Consequently, under OBC, IS will be preserved in S.C., but is broken in As.C.s. 
In fact, L-As.C. and R-As.C. are mirrors of each other since $P_0 H_{\text{LA}}^{\text{}}(k)P_0^\dagger$=$H_{\text{RA}}(-k)$, 
while S.C. is the mirror of its own.

Additionally, we find that the Bloch Hamiltonian of S.C., after some rearrangement, can be persymmetric, i.e. symmetric with respect to the anti-diagonal line, 
  $\tilde{H}_{\text{SC}}=J_3\tilde{H}_{\text{SC}}^{T}J_3$. 
Here 
\begin{equation*}
J_3=\begin{pmatrix}
0 & 0 &1 \\
0& 1& 0 \\
1& 0 & 0
\end{pmatrix},
\end{equation*}
is the exchange matrix, and 
\begin{equation}
\tilde{H}_{\text{SC}} =\left(\begin{array}{ccc}
0 & \Omega & -t_1-t_1e^{-i k} \\
\Omega & -2 t_{2} \cos k & \Omega  \\
-t_1-t_1e^{i k} & \Omega & 0
\end{array}\right),
\end{equation}
is the Bloch Hamiltonian of S.C. when the order of the sulattice basis is changed from $\{A, B, C\}$ to $\{A, C, B\}$. 
Note that the persymmetry is not host for As.C.s.
The persymmetry has been studied in certain quantum systems~\cite{huckle2013exploiting} but is less explored in topological matters. 
A recent work~\cite{palumbo2023topological} found that it may relate to indirect bandgaps  in certain systems.  We will also discuss the indirect bandgaps later.

In order to show IS more clearly, we also discuss it in an OBC terminated system in real space,
\begin{equation}
P=J_{N} \otimes P_0.
\end{equation}
For the real space Hamiltonian of a terminated system with $N=3M_t$ sites, we have $P H_{\text{SC}}P^\dagger =H_{\text{SC}}$ and $P
H_{\text{LA}}P^\dagger=H_{\text{RA}}$. Therefore, we refer to S.C. as the inversion-symmetric case and As.C.s as the IS broken cases.

Because L-As.C. and R-As.C. can be connected through the unitary operator $P$, for simplicity, we only discuss S.C. (symmetric case) and L-As.C. (IS broken case) in the following.



\subsection{Band structure} \label{band}

Since $H_{\text{SC}}(k)$, $H_{\text{LA}}(k)$, and $H_{\text{RA}}(k)$ can be transformed into each other through unitary transformations under PBC, they possess identical band structures.

Our investigation delves into the consequences of energy gap closures resulting from variations in the real-valued parameters, $\Omega$, $t_1$, and $t_2$. 
We present a comprehensive phase diagram illustrating diverse band closures across different parameter regions, accompanied by representative band structures (refer to Fig. \ref{band structure}).

Similar to the SSH model, the uneven ladder features direct bandgap closure as shown by the blue line in Fig. \ref{band structure}(a), where band indices $c_1$ and $c_2$ exist such that $E_{c_1}(k)=E_{c_2}(k)$, with $c_1\neq c_2$. However, unlike the SSH model, due to the lack of chiral symmetry in our model, the indirect bandgap closures occur as shown by the red and black lines in Fig. \ref{band structure}(a), where band indices $c_1$ and $c_2$ exist such that $E_{c_1}(k)=E_{c_2}(k^{\prime})$, with $c_1\neq c_2$ and $k\neq k^{\prime}$. 
Additionally, there is a trivial bandgap closure scenario when $\Omega=0$, in which case the system reverts to two independently periodic simple chains. The overlap of the band structures of the two single chains within the smaller first Brillouin zone leads to a trivial bandgap closure.

As denoted by the blue line in Fig. \ref{band structure}(a), the bandgap closure occurs at
\begin{equation}
\Omega=\pm 2 t_1 \sqrt{1+\frac{t_2}{t_1}}, \label{blueline}
\end{equation}
where $t_2/t_1 \ge -1$. 
Notably, the gap closes and reopens as the parameter crosses the blue line. 
Notice that if $t_1>0$ ($t_1<0$), the direct bandgap is located in between the second band and the third band (the first band and the second band).

The red line in Fig. \ref{band structure}(a) corresponds to an indirect gap closure or opening at
\begin{equation}
\Omega=\pm \frac{2t_1}{t_1/t_2+2} \sqrt{1+\frac{t_2}{t_1}},
\end{equation}
with $t_2/t_1 \ge 0$. 
Bands within the parameter region at the right side of  the red line experience a global gap closure and remain closed. 
Notice that if $t_1>0$ ($t_1<0$), this indirect bandgap is located between the first band and the second band (the second band and the third band).

The thick black line in Fig. \ref{band structure}(a) represents another indirect gap closure or opening at
\begin{equation}
\Omega=\begin{cases}
 \pm \sqrt{2}t_1\sqrt{1-\frac{t_2}{t_1}} ,\quad  \frac{t_2}{t_1}\le -\frac{1}{3}
\\\pm \frac{2t_1}{t_1/t_2+2}\sqrt{1+\frac{t_2}{t_1}} ,\quad  -\frac{1}{3}<\frac{t_2}{t_1}\le0
\end{cases}    .
\end{equation}
This indicates that if the parameters are located at the left side of the thick black line, the bands will feature a global gap closure, and will not be opened again. 
Notice that if $t_1>0$ ($t_1<0$), this indirect bandgap is located between the second band and the third band (the first band and the second band).

We also note that when $t_2=0$, such an uneven ladder can be reduced to a triangular-like ladder~\cite{Li_2023} or a chain with ring structure~\cite{velasco2019classification}. 
Specifically, if $\Omega/t_1=\sqrt{2}$ is also satisfied, there will be a flat band, i.e. a band whose energy is independent of $k$, see Fig. \ref{band structure} (b12).

\section{Projecting to bipartite chain\label{projection}} 

The proposed simplest uneven ladder establishes a strong connection with the SSH model~\cite{Su_1979}, whose Bloch Hamiltonian reads
\begin{equation}
H_{\text{SSH}}(k)=\begin{pmatrix}
0 & -v-w e^{-ik} \\
 -v-we^{ik}&0
\end{pmatrix}, \label{SSH}    
\end{equation}
where $v$ and $w$ denote the constant inter- and intra-cellular hopping strength. 
The case $|v| < |w|$ corresponds to a topological phase, while $|v| > |w|$ corresponds to a topology-trivial phase. 
In the following we demonstrate the relationship between the simplest uneven ladder and SSH model through a projection approach. 

We first focus on S.C..  
The Bloch Hamiltonian of S.C. can be decomposed as three parts: $H_{\text{SC}}(k)=H_{AB}+H_{C}+H_{\Omega}$, where 
\begin{eqnarray}
H_{AB}&=&\begin{pmatrix}
 0  & \Gamma  & 0\\
 \Gamma^* & 0  & 0\\\nonumber
  0& 0 &0
\end{pmatrix},
H_C=\begin{pmatrix}
  0&0  &0 \\
  0& 0 & 0\\
  0& 0 &\delta 
\end{pmatrix},
H_{\Omega} =\begin{pmatrix}
 0 & 0 &\Omega \\
  0& 0 &\Omega \\
 \Omega & \Omega &0
\end{pmatrix}.
\end{eqnarray}
Here $\Gamma=-t_1-t_1 e^{-ik}$, $\delta=-2t_2\cos k$. 
We project $H_1(k)$ to the sub-space spanned by $A$ and $B$ sites of chain-$1$~(for details, see  Appendix \ref{app_proj}), 
and obtain a $2\times2$ projected effective Hamiltonian,
\begin{eqnarray}
H_{\text{SC,eff}}^{(n)}(k)=&H_{AB}+\frac{\Omega^2}{E^{(n)}(k)-\delta }\begin{pmatrix}
 1 &1 \\
 1 &1
\end{pmatrix},
\end{eqnarray}
which can be further written as
\begin{eqnarray}
\begin{aligned}
H_{\text{SC,eff}}^{(n)}(k)=&\frac{\Omega^2}{E^{(n)}(k)+2t_2 \cos k }I  \\&+ \begin{pmatrix}
0 & -v-w e^{-ik} \\
 -v-we^{ik}&0
\end{pmatrix}. \label{eff1}    
\end{aligned}
\end{eqnarray}
Here $v=t_1-\Omega^2/[E^{(n)}(k)+2t_2\cos k)] $,  $w=t_1$. 
The first term in (\ref{eff1}) can be viewed an overall energy shift; 
the second term resembles the Bloch Hamiltonian of SSH model, (\ref{SSH}). 
Thus, we can regard $H_{\text{SC,eff}}^{(n)}(k)$ as a generalized  SSH model. 
The generalization involves the dependence of the intercellular hopping coefficient $v$ on (original) band index $n$ and momentum $k$. 
Since $H_{\text{SC,eff}}^{(n)}$ lacks diagonal terms, it possesses chiral symmetry. 
it is worth noting that, the existence of the mapping to SSH model is ensured by the mutual symmetry between $A$ and $B$ sites (the exchange of $A$ and $B$ sites preserves the system's properties). 
When on-site potential difference term of the $A$ and $B$ sublattice or the intensity difference of the $C$ sublattice coupling to the $A$ and $B$ sublattice are introduced, the effective projected Hamiltonians of this two types will inevitably include $\sigma_z$ term, thus corresponding to the RM model instead of SSH model.
In the following we discuss the topological property of the generalized SSH model.
\begin{figure}[t]
\includegraphics[width=8.6cm]{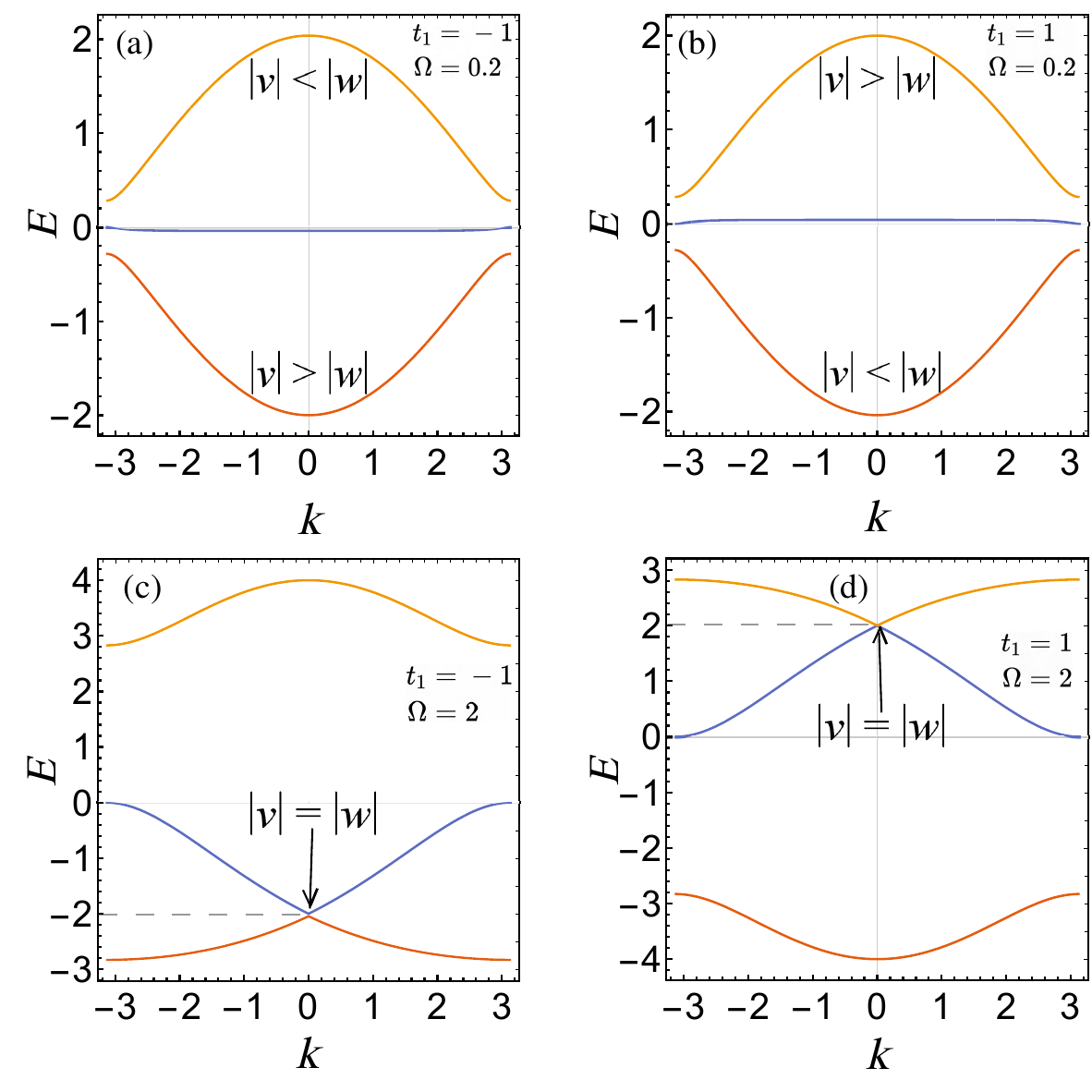}
\caption{Band structure of the original Hamiltonian and comparison of intracellular and intercellular hopping strength of the effective projected Hamiltonian at $t_2=0$ in S.C..}   \label{backssh1}
\end{figure}

We first set $|\Omega|$  sufficiently larger than $t_1$, and set $t_2=0$. The band energies now are approximately equal to $E^{(1)}=-\sqrt{2}\Omega$, $E^{(2)}=0$, $E^{(3)}=\sqrt{2}\Omega$, so that $v=t_1+\Omega/\sqrt{2}$, $w=t_1$ for the first band, and $v=t_1-\Omega/\sqrt{2}$, $w=t_1$ for the third band. 
In this case, the first band and the third band in S.C. correspond to the SSH model with topological trivial phase since $|v| > |w|$.

Then, we assume $|\Omega|$ is sufficiently smaller than $t_1$, and set $t_2=0$. In this way, the band energies are approximately equal to $E^{(1)}=2t_1\cos (k/2)$, $E^{(2)}=0$, $E^{(3)}=-2t_1\cos (k/2)$. For the first band, $|v|>|w|$ ($|v|<|w|$) since $E^{(1)}>0$ ($E^{(1)}<0$) for $t_1>0$ ($t_1<0$). On the other hand, for the third band, $|v|<|w|$ ($|v|>|w|$) since $E^{(3)}<0$ ($E^{(3)}>0$) for $t_1>0$ ($t_1<0$). 
Consequently, for sufficiently small $t_2$ and $\Omega$, this results in that, when $t_1<0$, the third band is topologically nontrivial and the first band is topologically trivial. On the other hand, when $t_1>0$, the third band is topologically trivial and the first band is topologically nontrivial, as Fig. \ref{backssh1}(a) and (b) show.

Next, we consider the case of $|v| = |w|$, so that it requires $v=-w$, i.e., $\Omega^{2}/E=2t_1$ for $t_2=0$. 
We expect it to be the same as the SSH model, with Dirac-like gap closing at $k=0$. In fact, when $t_2=0$, $|\Omega|=2|t_1|$ and $t>0$ ($t<0$), $E^{(1,2)}(k=0)=+(-)2t_1$, the Dirac-like gap closing does exist. Besides, they are indeed satisfied with this relationship $\Omega^{2}/E=2t_1$, i.e, $|v| = |w|$ (refer to Fig. \ref{backssh1}(c) and (d)].

\begin{figure}[t]
\includegraphics[width=8.6cm]{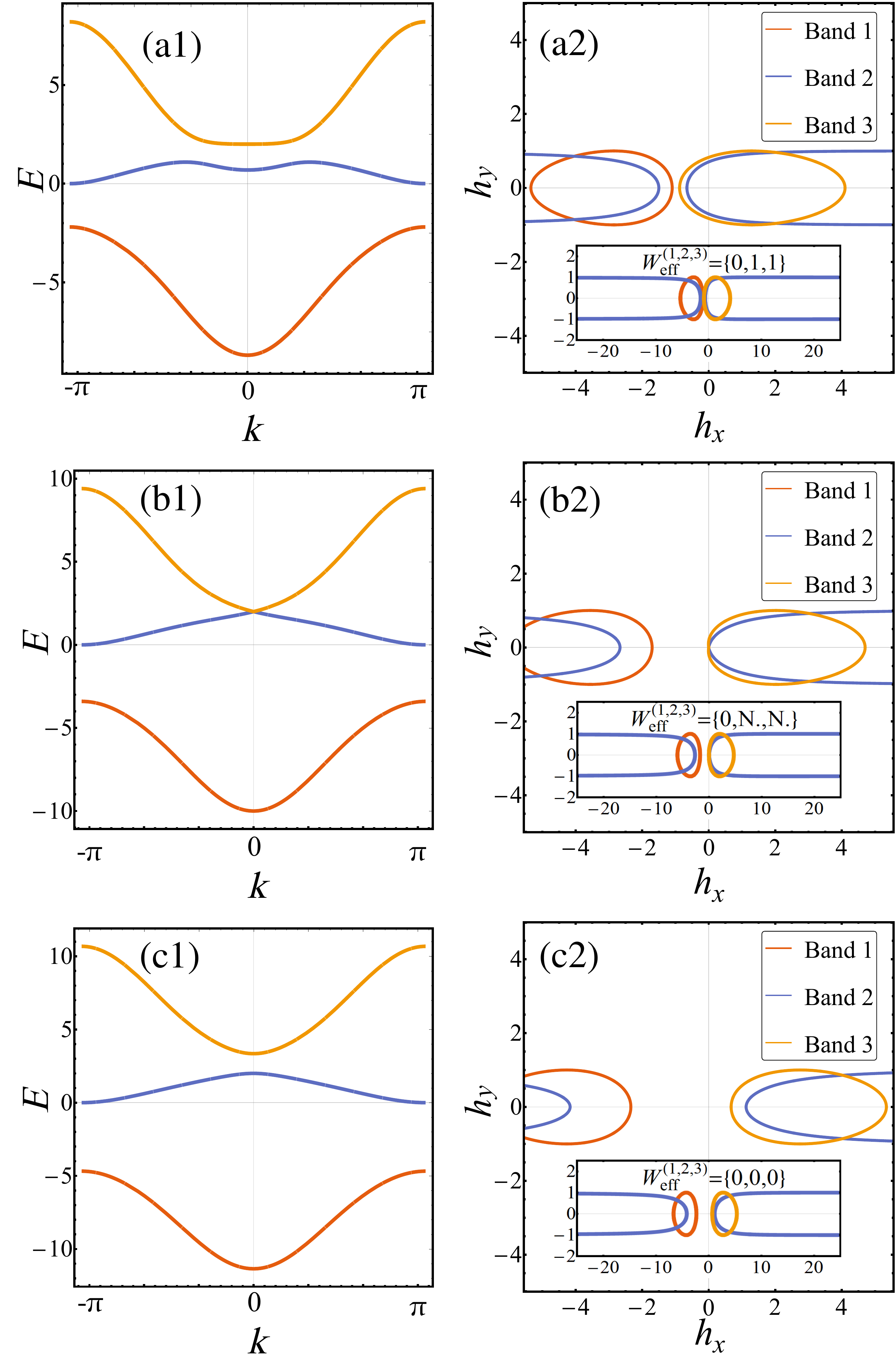}
\caption{Effective winding numbers of S.C.. (a1), (b1), (c1) are band structures, with corresponding (a2), (b2), (c2) illustrating trajectories of $\{\boldsymbol{h}^{(n)}\}$ and $\{W_{\text{eff}}^{(n)}\}$ for each band, shown at $\Omega=3$, $4$, $5$, respectively. Here, $t_2=3$, $t_1=1$. } \label{3000}
\end{figure}

Now let us think about how to figure out whether system is in the topological phase for arbitrary $t_1$, $t_2$, $\Omega$. 
Although we can always map our model to the SSH model no matter how the parameters are chosen, it is complicated to directly compare $|v|$ and $|w|$ since the involved band energy $E^{(n)}$ needs to be calculated numerically. 
Therefore we aspire to solve this problem through the winding number of the effective SSH model.
In fact, winding number is not defined for a general $3\times3$ Hamiltonian, but we can define it for the $2\times2$ effective projected Hamiltonian, which can be decomposed in terms of Pauli matrices, $\tilde{H}_{\text{SC,eff}}^{(n)}=\boldsymbol{h}^{(n)} \hat{\sigma}=h^{(n)}_x \sigma_x+h^{(n)}_y \sigma_y$. 
 Specifically, the expressions of $h^{(n)}_x$ and $h^{(n)}_y$ in S.C. are 
\begin{eqnarray}
    h^{(n)}_x(k)&=&-v-w \cos k \nonumber\\
    &=&-t_1(1+\cos k)+\frac{\Omega^2}{E^{(n)}(k)+2t_2 \cos k} ,\\
    h^{(n)}_y(k)&=&-w \sin k=-t_1 \sin k.
\end{eqnarray} 
In this way, the key point is the trajectory of the effective Hamiltonian on the $h_x-h_y$ plane as momentum $k$ varies continuously through the first Brillouin zone from 0 to $2\pi$. 
 Just like the SSH model, when the trajectory surrounds the origin, the system is in the topological phase; the opposite is true for the trivial phase.
The winding number of the effective projected Hamiltonian (dubbed the effective winding number) can be written as 
\begin{eqnarray}
    W_{\text{eff}}^{(n)}&=&\frac{1}{2\pi}\int_{-\pi}^{\pi} dk[\tilde{\boldsymbol{h}}^{(n)}(k) \times \partial_{k} \tilde{\boldsymbol{h}}^{(n)}(k)]_z \nonumber \\
    &=&\frac{1}{\pi} \int_{-\pi}^{\pi} dk \tilde{h}_x^{(n)}(k) \partial_{k} \tilde{h}_y^{(n)}(k), \label{winding}
\end{eqnarray}
where  $\tilde{\boldsymbol{h}}(k)=\boldsymbol{h}(k)/|\boldsymbol{h}(k)|$. 
It is worth noting that each energy band may independently correspond to an effective winding number as Fig. \ref{3000} shows.
Note that, although the trajectory of the second band extends infinitely at both ends and is not closed, positive infinity and negative infinity are connected together. 

\begin{figure}[t]
\includegraphics[width=8.6cm]{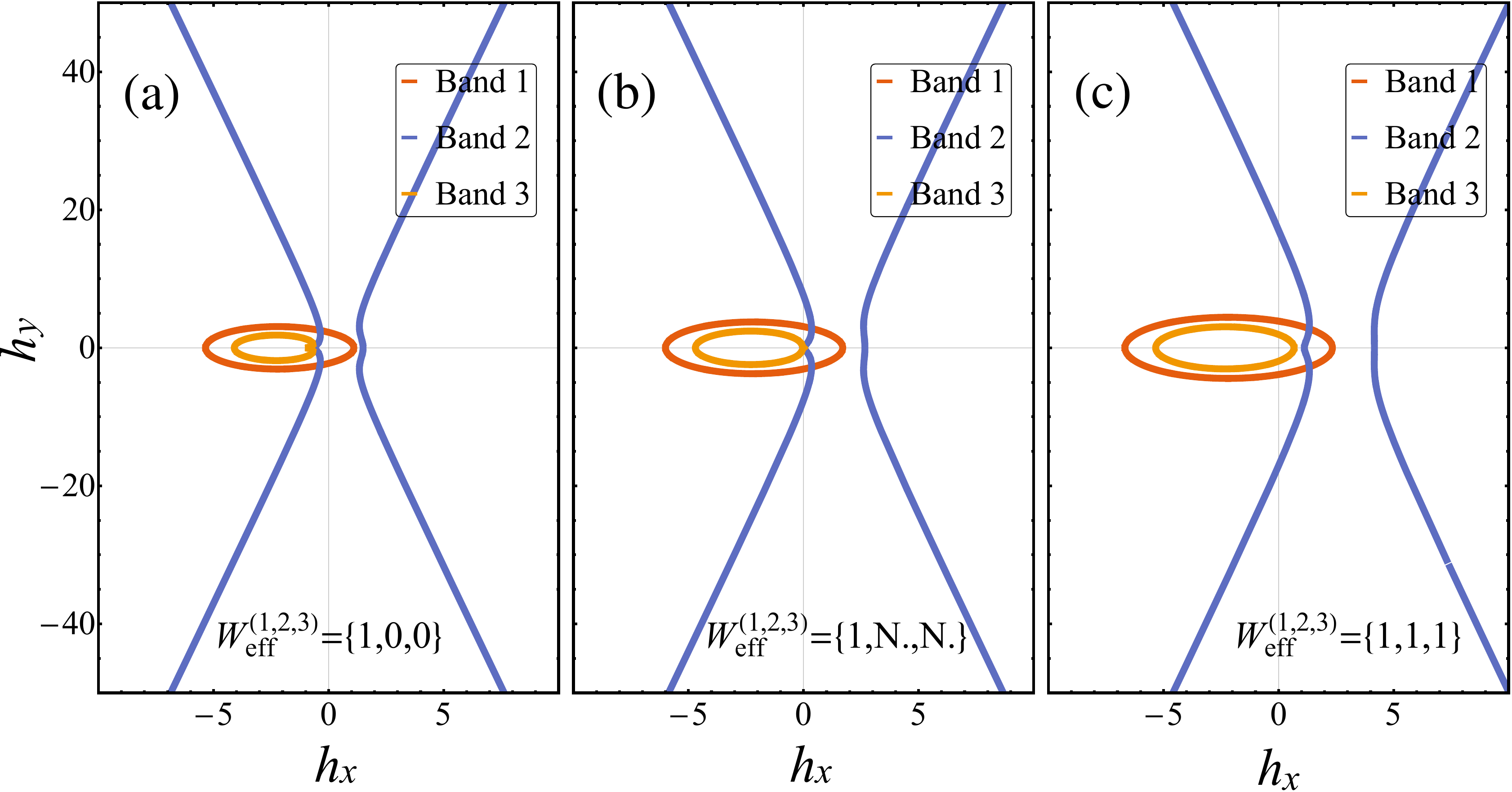}
\caption{Effective winding numbers of L-As.C.. (a), (b), (c) illustrate trajectories of $\{\boldsymbol{h}^{(n)}\}$ and $\{W_{\text{eff}}^{(n)}\}$ for each band, shown at $\Omega=3$, $4$, $5$, respectively. Here, $t_2=3$, $t_1=1$.} \label{asw}
\end{figure}

The introduction of the effective winding number is both convenient and efficient, since the winding number is determined by the energy band rather the Bloch wavefunction. 
We shall also note that such an approach is not limited to the uneven ladder studied here. It provides a shortcut for analyzing the topology of generally tripartite models (refer to Appendix \ref{app_proj}). 
It is in principle applicable to be extended to more intricate lattice models.

In this section, we have only discussed the S.C. so far. Now, let us discuss  projected effective Hamiltonian $H_\text{LA,eff}^{(n)}$ for L-As.C.. $H_{\text{LA,eff}}^{(n)}$ will remain the form of (\ref{eff1}), 
while the effective hopping become 
$v=t_1$, $w=t_1-\Omega^2/[E^{(n)}(k)+2t_2 \cos k]$. 
We can calculate the effective winding number as well, as shown in  Fig. \ref{asw}. 

\section{Zak phase}\label{Zak}

In 1D non-interacting fermionic systems with either chiral symmetry or IS, the appearance of topological phases is safeguarded by a topological invariant known as the Zak phase \cite{zak1989berry}. 
Recent works reveal that the Zak phase can be decomposed as intercellular Zak phase and intracellular Zak phase~\cite{kudin2007berry, Rhim2017}.
The former, contingent upon space coordinates selection, serving as a $\mathbb{Z}_2$ topological invariant, categorizes topological phases aligning with bulk-boundary correspondence, while the latter exhibits independence. 
They also correspond to the dipole moment and extra charge accumulation, respectively.
In this section, we explore the two types of the Zak phases. 
As introduced in Sec.~\ref{model}, there can be two types of configurations, based on the choice of unit cell or cut on the edge—symmetric and asymmetric configuration. 
Under periodic boundary conditions, these configurations prove to be essentially identical up to a unitary transformation. The symmetric configuration, characterized by IS, falls within the $\mathbb{Z}_2$ classification concerning the intercellular Zak phase.

\subsection{Intracellular Zak phase vs intercellular Zak phase\label{ii}}

\begin{figure}[t]
\includegraphics[width=8.6cm]{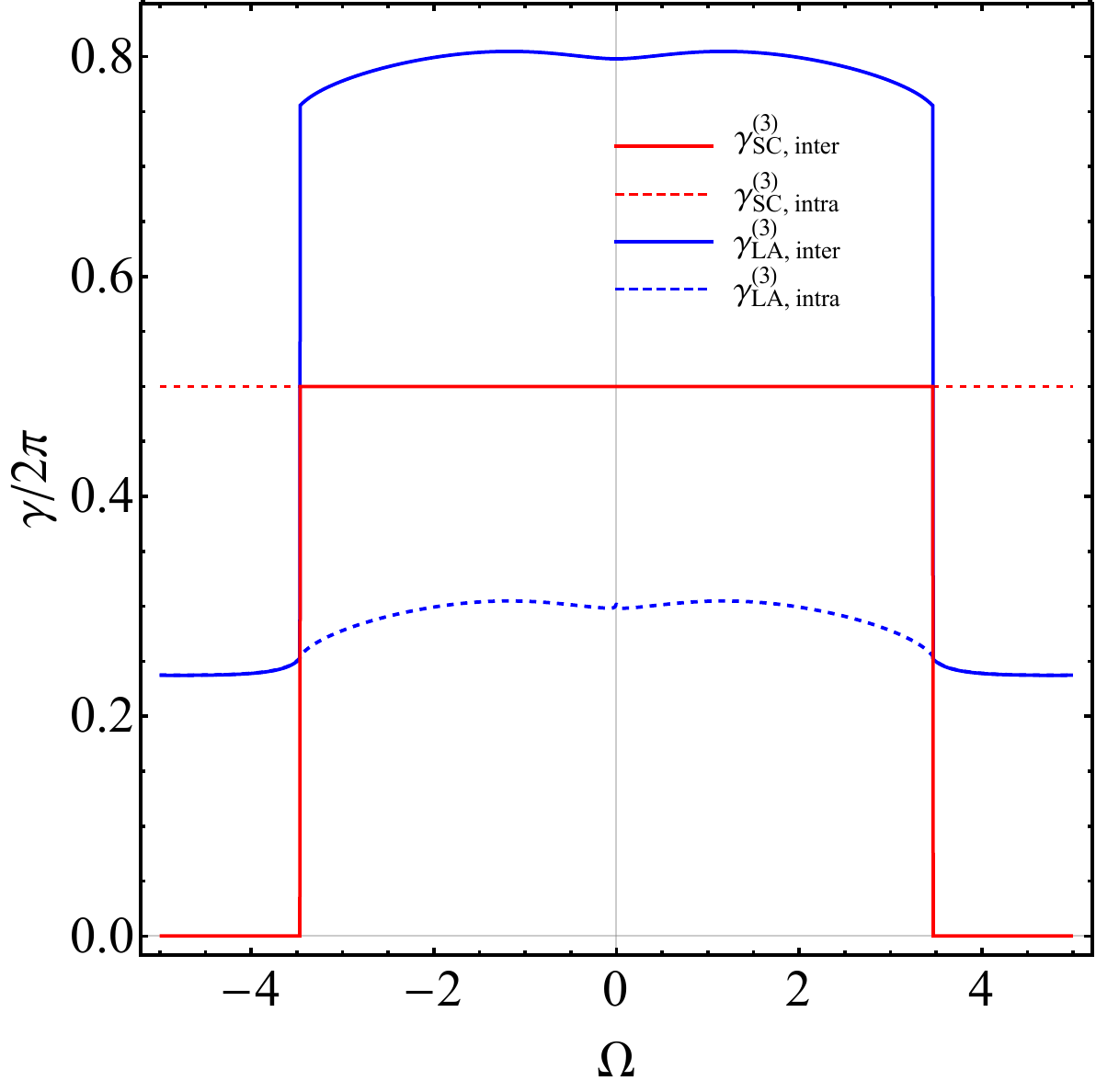}
\caption{Intercellular Zak phase and intracelluar Zak phase of third band of S.C. and L-As.C. at $t_1=1$, $t_2=2$.}   \label{3}
\end{figure}

Zak phase of the \( n \)th band is defined as
\begin{equation}
    \gamma^{(n)} = i \int_{0}^{2 \pi} dk \left\langle u_{n, k} \middle|\partial_{k}\middle| u_{n, k} \right\rangle.
\end{equation}
The inner product here can be explicitly written as 
\begin{equation}
    \left\langle u_{n, k} \middle|\partial_{k}\middle| u_{n, k} \right\rangle = \int_{\Omega_{m^{\prime}}} dx \ u_{n, k}^{*}(x) \partial_{k} u_{n, k}(x),
\end{equation}
where $ u_{n, k}(x) = \sqrt{N} e^{-i k x} \psi_{n, k}(x) $, the periodic part of the Bloch function. $ \Omega_{m^{\prime}} $ means the $m^{\prime}$-th unit cell.
Within the tight-binding approximation, we use a set of orthogonal wave functions localized on sites as the basis to expand the Bloch function
\begin{equation}
    \psi_{n, k}(x) = \frac{1}{\sqrt{N}} \sum_{m}^{M_t} \sum_{l}^{L_t}e^{i k m} \alpha_{n, k}^{l} \phi_{}^{m,l}(x),
\end{equation}
where \( \phi_{}^{m,l}(x) \) is localized at the \( l \)-th intercelluar site of the \( m \)-th cell. The coefficients \( \alpha_{n, k}^{l} \) are nothing but the solutions of the tight-binding Hamiltonian.
Zak phase may change according to the gauge of coordinates of sites (or we can say real-space origin). One can divide the Zak phase into intracellular Zak phase and intercellular Zak phase as following \cite{kudin2007berry, Rhim2017},
\begin{eqnarray}
    \gamma^{(n)} &=& \gamma^{(n)}_{\text {intra}} + \gamma^{(n)}_{\text {inter}},\\
    \gamma^{(n)}_{\text {intra}} & = & N \int_{0}^{2\pi} dk \int_{\Omega_{m^{\prime}}} dx \ x \left|\psi_{n, k}(x)\right|^{2} - 2 \pi m^{\prime},\\
    \gamma^{(n)}_{\text {inter}} &=& i \sum_{m=1}^{M_t} \sum_{l}^{L_t} \int_{\mathrm{BZ}} dk \ \alpha_{n,k}^{l} \frac{\partial}{\partial k} \alpha_{n,k}^{l}.
\end{eqnarray}
where the intercellular Zak phase is independent of the real-space origin. 
For the simplest uneven ladder, we have
\begin{equation}
    H(k)\begin{pmatrix}
 \alpha ^{A}_{n,k}\\\alpha ^{B}_{n,k}
 \\\alpha ^{C}_{n,k}

\end{pmatrix}
=
E_{n,k}\begin{pmatrix}
 \alpha ^{A}_{n,k}\\\alpha ^{B}_{n,k}
 \\\alpha ^{C}_{n,k}

\end{pmatrix}.
\end{equation}
The Bloch wave function will be 
\begin{eqnarray}
\begin{aligned}
\psi_{n,k}(x)=&\frac{1}{\sqrt{N}}\sum_{m}^{M_t} e^{ikm} [\alpha_{n,k}^{A} \phi^{m,A} (x)
\\
&+\alpha_{n,k}^{B}(x) \phi^{m,B}+\alpha_{n,k}^{C} \phi^{m,C}(x)],     
\end{aligned}   
\end{eqnarray}
where  $\phi^{m,A}(x)$ is localized at $x=m+1/4$, $\phi^{m,B}(x)$ is localized at $x=m+3/4$, $\phi^{m,C}(x)$ is localized at $x=m+1/2$ ($x=m$) for S.C. (As.C.).  In this way, we can calculate intracellular and intercellular Zak phase under PBC as Fig. \ref{3} shows.

Actually, only the intercellular Zak phase is the topological invariant that satisfies the bulk-boundary correspondence \cite{Rhim2017}. As TABLE \ref{zakwinding} shows, we compare the intercellular Zak phase with the winding number of the effective projected Hamiltonian in both S.C. and L-As.C., where I, II, III, IV, V donate the parameters region in Fig. \ref{band structure}(a). Notice that in S.C., there is correspondence between intercelluar Zak phase and effective winding number $W_{\text{SC,eff}}^{(1,2,3)}/\pi=\gamma^{(1,2,3)}_{\text{SC,inter}}$; on the other hand, there is no such correspondence in L-As.C.. 

\begin{table}[t]
    \centering
    \caption{the comparison of intercellular Zak phase and effective winding number ($t_1>0$).}
    \label{zakwinding}
    \begin{tabular}{cccccc}
    \toprule
\textbf{} & \textbf{I/V} & \textbf{II/III/IV}  \\ \midrule
\midrule
 & $\gamma^{(1,2,3)}_{\text{inter}}/\pi$=\{0, 0, 0\} & $\gamma^{(1,2,3)}_{\text{inter}}/\pi$=\{$1$, $1$, 0\} \\ 
\textbf{S.C.} & ~  \\ 
 & $W_{\text{eff}}^{(1,2,3)}$=\{0, 0, 0\} & $W_{\text{eff}}^{(1,2,3)}$=\{1, 1, 0\}  \\ 
 \midrule
 & $\gamma^{(1,2,3)}_{\text{inter}}/\pi$=fractions & $\gamma^{(1,2,3)}_{\text{inter}}/\pi$=fractions  \\ 
 \textbf{L-As.C.} & ~  \\ 
 & $W_{\text{eff}}^{(1,2,3)}$=\{1, 1, 1\} & $W_{\text{eff}}^{(1,2,3)}$=\{0, 0, 1\}   \\  \bottomrule
    \end{tabular}    \label{zakandwT}
\end{table}

We can comprehend the breakdown of the correspondence between the intercelluar Zak phase and the effective winding number in L-As.C. from the perspective of symmetry protection: On the one hand, original Hamiltonian of L-As.C. lacks chiral symmetry and spatial inversion symmetry, thus leading to the loss of quantization properties in the Zak phase. On the other hand, the effective projected Hamiltonians $\tilde{H}_{\text{LA,eff}}^{(1)}(k), \tilde{H}_{\text{LA,eff}}^{(2)}(k), \tilde{H}_{\text{LA,eff}}^{(3)}(k)$ of L-As.C. are protected by chiral symmetry, hence maintaining quantized effective winding numbers $W_{\text{SC,eff}}^{(1,2,3)}$.

\subsection{Physical counterparts}
\begin{figure}[!h]
\includegraphics[width=8.6cm]{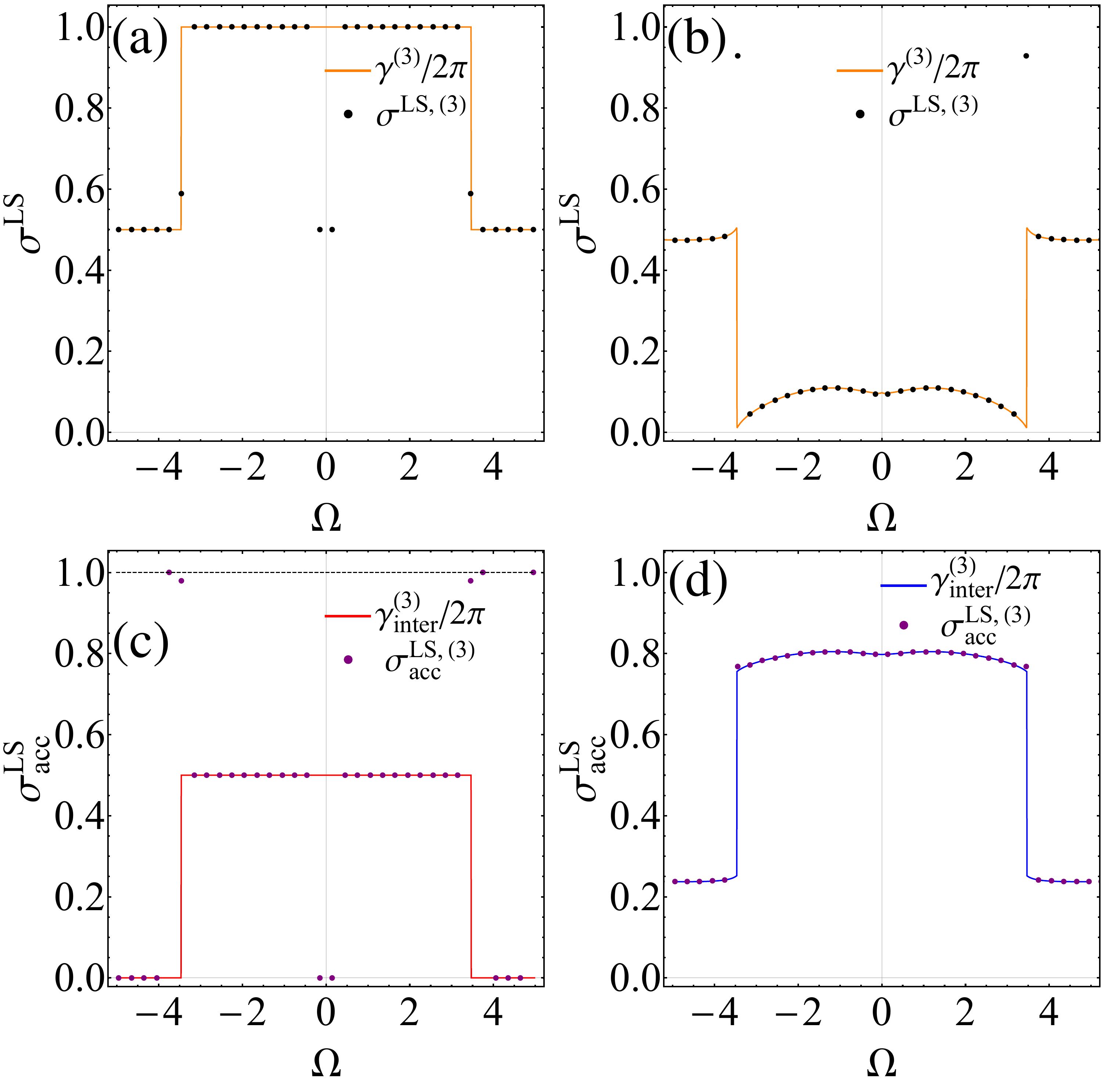}
\caption{
(a-b) Comparison between the (total) Zak phase (mod $2\pi$) and the bound surface charge (mod 1) at the left edge of OBC terminated coupled chain for third band in S.C. and L-As.C., respectively. 
(c-d) Comparison between the intercelluar Zak phase (mod $2\pi$) and the extra charge accumulation (mod 1) at the left edge of OBC terminated coupled chain in S.C. and L-As.C., respectively. 
The length of the terminated coupled chain we took is 360 unit cells for calculating bound surface charge and extra charge accumulation.  
Note that due to size effects, there are discrepancies in the correspondence near the direct bandgap closure for $\Omega \to 0$.
We set $t_1=1$, $t_2=2$, $q=1$ here. } \label{4}
\end{figure}
 The physical counterpart of $\gamma^{(n)}$ is electric polarization $\boldsymbol{P}$ \cite{vanderbilt1993electric}, so that it is equivalent to the corresponding to $\sigma^{\text{LS}}$ since we have the relationship $\sigma^{\text{LS}}=\boldsymbol{P}\hat{n}$, where $\sigma^{\text{LS}}$ is the surface bound charge of left side boundary and we set $\hat{n}=+(-)\hat{x}$ as the surface orientation for left (right) edge in 1D terminated (OBC) system\cite{Rhim2017}. 
 The physical interpretations of intracellular and intercellular Zak phase also have been given, that is, classical surface bound charge $\sigma_{cl}$ and extra charge accumulation $\sigma_{acc}$ of OBC system, respectively \cite{Rhim2017, vanderbilt1993electric}.  
 it is a strong correspondence since one can use the Zak phase calculated under PBC to know the boundary charge physical quantities of terminated system under OBC. 
 More clearly, when we consider the left edge, they are defined as \cite{Rhim2017, vanderbilt1993electric, baldereschi1988band, kudin2007berry}
 \begin{eqnarray}
     \sigma^{\text{LS}}&=&\sigma^{\text{LS}}_{cl}+\sigma^{\text{LS}}_{acc}
     \nonumber \\&=& \int_{x_{\ell_{}}}^{x_{\ell_{}}+1} d x \  x \rho_{\text{ts}}(x)+\int_{x_0}^{x_{\ell_{}}} d x \ \rho_{\text{ts}}(x),
 \end{eqnarray}
where $\sigma_{cl}^{\text{LS}}$ and $\sigma_{acc}^{\text{LS}}$ respectively are the two terms of the above formula, $x_0$ is the coordinate of the first site from the left, $x_{\ell_{}}$ is the coordinate of the site far from boundaries (as long as $x_{\ell}$ is selected sufficiently far away from the boundaries, it will not affect the result) and $\rho_{\text{ts}}(x)=q \boldsymbol{\psi}^{\text {ts}}(x)^{\dagger} \boldsymbol{\psi}^{\text {ts}}(x)$ is the charge density distribution of terminated system which can be obtained by solving the
eigenvectors $\boldsymbol{\psi}^{\text {ts}}(x)=\left\{\psi_{i}^{\text {ts}}(x), \ldots, \psi_{i^\prime}^{\text {ts}}(x)\right\}^{\mathrm{T}}$ of OBC Hamiltonian in real-space \cite{Rhim2017}, where $\{i...i^\prime\}$ are indexes of OBC eigenvectors we are interested in, and $q$ is the charge of a particle. In the example of tight binding OBC system with size $3M_t$, if we are only interested in the $n$-th band (only consider occupied $n$-th band, $n=1, 2, 3$), it can be $\boldsymbol{\psi}^{\text {ts,(n)}}(j)=\left\{\psi_{(n-1)M_t+1}^{\text {ts}}(j), \ldots, \psi_{nM_t}^{\text {ts}}(j)\right\}^{\mathrm{T}}$, then we can calculate $\rho_{\text{ts}}^{(n)}(j)$ and $\sigma^{\text{LS},(n)}$, where $j$ donates the $j$-th site.

For the uneven ladder, we checked the relationship between Zak phase and surface bound charge, and the relationship between intercellular Zak phase and extra charge accumulation for third band as Fig. \ref{4} shows.  We set lattice constant $a=1$,  $q=1$, and let $\gamma^{(n)}$, $\gamma^{(n)}_{\text {intra}}$, $\gamma^{(n)}_{\text {inter}}$ modulo $2\pi$, and let  $\sigma^{\text{LS}}$, $\sigma^{\text{LS}}_{cl}$, $\sigma^{\text{LS}}_{acc}$ module 1. it is worth noting that, when the parameters are not very close to the direct band closing points, they correspond well with each other.
We did not demonstrate the right edge, since the system as a whole appears neutral, and the right side should differ from the left side only by a negative sign.

\subsection{Relation between S.C.’s and L-As.C.’s intercelluar Zak phases\label{zakrela}}

Note that, although the Hamiltonians of S.C. and L-As.C. under PBC are related by unitary transformations, their intercellular Zak phases differ. 
On the one hand, the intercellular Zak phase $\gamma^{(n)}_{\text{SC},\text {inter}}$ of S.C., possessing IS, is constrained to integer multiples of $\pi$. 
On the other hand, the intercellular Zak phase $\gamma^{(n)}_{\text{LA},\text {inter}}$ of L-As.C., lacking IS, takes fractional multiples of $\pi$ and lacks quantization properties. 
Indeed, we can analytically demonstrate this point. 
Next, we will discuss in detail the impact of unitary transformations between S.C. and L-As.C. on their intercellular Zak phases.

\begin{figure}[t]
\includegraphics[width=8.6cm]{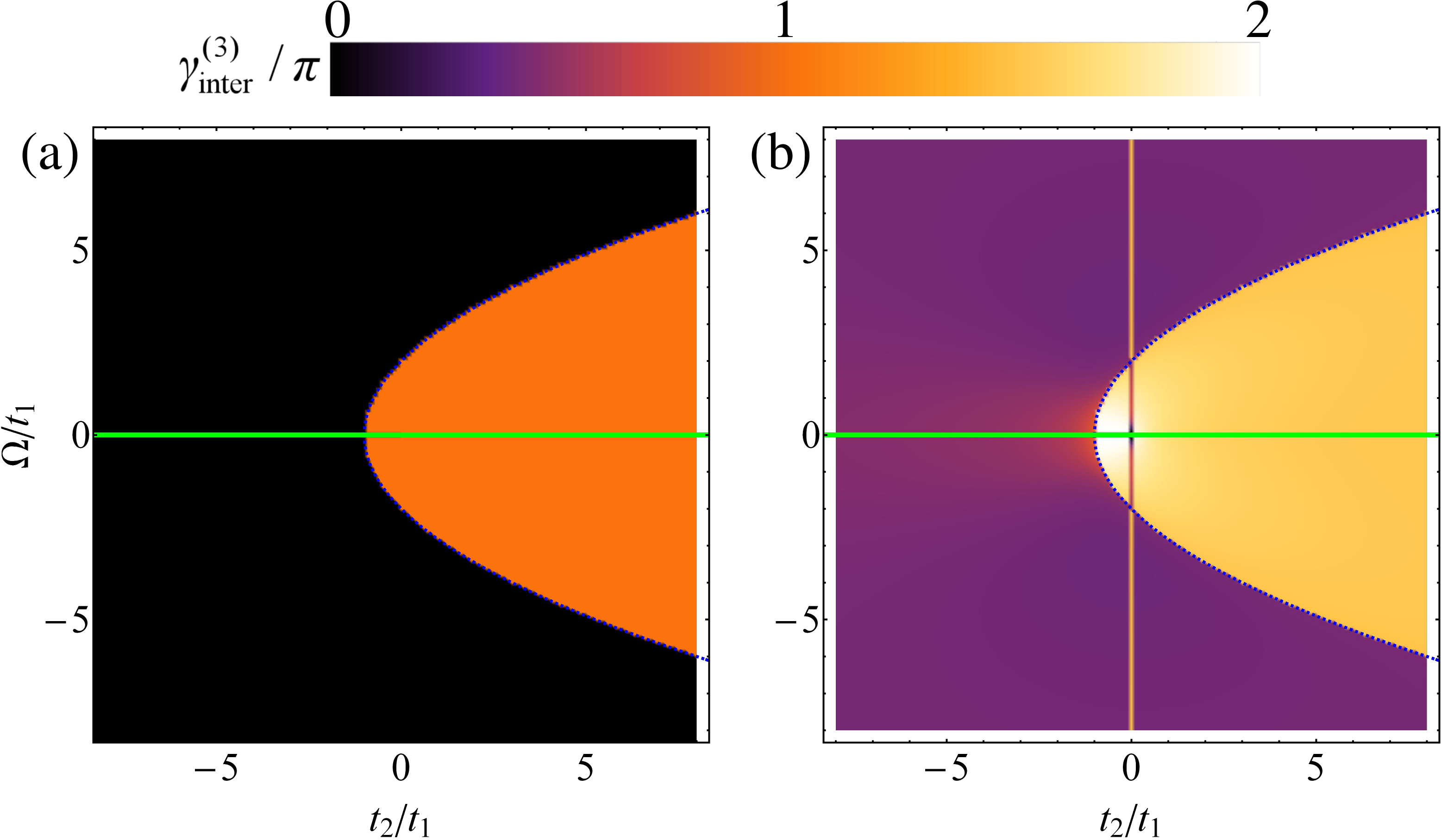}
\caption{
Intercellular Zak phases of S.C. and L-As.C. in the entire parameter space. (a) and (b) depict the intercelluar Zak phases of the third band, $\gamma_{\text{inter}}^{(3)}$ , for S.C. and L-As.C. respectively. The blue dashed lines represent parameters corresponding to direct bandgap closure. The green line corresponds to $\Omega=0$, where there is no coupling resulting in band overlap between the upper and lower chains, leading to an ill-defined (Null) intercelluar Zak phases.
} \label{interzak}
\end{figure}

Again, we write down the connection between $H_{\text{SC}}$ and $H_{\text{LA}}$:
\begin{equation}
    U(k) H_{\text{SC}}(k) U^{\dagger}(k)=H_{\text{LA}}(-k). \label{u12}
\end{equation}
First, due to the similarity in the band structures of S.C. and L-As.C., we may conveniently establish the following eigenvalue equation,
\begin{eqnarray}
    H_{\text{SC}}(k)\left |  \psi_{g_1}  \right \rangle =E_g(k)\left |  \psi_{g_1}  \right \rangle, \label{eigen1}\\
    H_{\text{LA}}(k)\left |  \psi_{g_2}  \right \rangle =E_g(k)\left |  \psi_{g_2}  \right \rangle, \label{eigen2}
\end{eqnarray}
where the eigenstates can be written as $\left |  \psi_{g_1}  \right \rangle=(g_1^{(1)}, g_1^{(2)}, g_1^{(3)})^T$, $\left |  \psi_{g_2}  \right \rangle=(g_2^{(1)}, g_2^{(2)}, g_2^{(3)})^T$. Note that, because S.C. possesses IS, the following relations hold: $E_g(k)=E_g(-k)$ and $g_1^{(u)}(k)=g_1^{(u)}(-k)$ (for $u=1, 2, 3$). Then from (\ref{u12}) and (\ref{eigen1}), we obtain
\begin{equation}
    H_{\text{LA}}(k) U(-k) \left |  \psi_{g_1}  \right \rangle =E_g(k) U(-k) \left |  \psi_{g_1}  \right \rangle. \label{lag1}
\end{equation} 
Under the assumption of disregarding energy degeneracy in the $k$-space (direct bandgap open), we obtain:
\begin{equation}
   \left |  \psi_{g_2}  \right \rangle=U(-k) \left |  \psi_{g_1}  \right \rangle=\begin{pmatrix}
g_1^{(1)}(k) \\e^{i k}g_1^{(2)}(k)
 \\g_1^{(3)}(k))
\end{pmatrix} . \label{con}
\end{equation}
We write down the intercelluar Zak phase of $H_{\text{SC}}$ and $H_{\text{LA}}$ as 
\begin{eqnarray}
    \begin{aligned}
    \gamma_{\text{SC},\text{inter}}=&i \int_{0}^{2 \pi} d k\left\langle \psi_{g1} \left|\partial_{k}\right|\psi_{g1} \right\rangle
      \\
     =&i \int_{0}^{2 \pi} d k (g_1^{*(1)} i\partial_{k} g_1^{(1)}
      \\
     &+{g_1^{*(2)}} i\partial_{k} g_1^{(2)}+{g_1^{*(3)}} i\partial_{k} g_1^{(3)}), \label{g1}    
    \end{aligned}
\end{eqnarray}
\begin{eqnarray}  
    \begin{aligned}
     \gamma_{\text{LA},\text{inter}}=&i \int_{0}^{2 \pi} d k\left\langle \psi_{g2} \left|\partial_{k}\right| \psi_{g2} \right\rangle  \\
     =&i \int_{0}^{2 \pi} d k (g_2^{*(1)} i\partial_{k} g_2^{(1)}
      \\
     &+g_2^{*(2)} i\partial_{k} g_2^{(2)}+g_2^{*(3)} i\partial_{k} g_2^{(3)})\label{g2}.   
    \end{aligned}
\end{eqnarray}
Notice that ${g_2^{*(1)}}(k) i\partial_{k}g_2^{(1)}(k)={g_1^{*(1)}}(k) i\partial_{k}g_1^{(1)}(k)$, ${g_2^{*(3)}}(k) i\partial_{k}g_2^{(3)}(k)={g_1^{*(3)}}(k) i\partial_{k}g_1^{(3)}(k)$, and
\begin{equation}
  {g_2^{*(2)}}(k) i\partial_{k}g_2^{(2)}(k)= g^{*(2)}_1(k) i\partial_{k} g_1^{(2)}(k)+|g_1^{(2)}(k)|^2 , \label{3t}    
\end{equation}
where Eq. (\ref{con}) is used.
From Eq. (\ref{g1}), (\ref{g2}) and (\ref{3t}), we finally obtain
\begin{eqnarray}
 \gamma_{\text{LA},\text{inter}}-\gamma_{\text{SC},\text{inter}}=i \int_{0}^{2 \pi} d k |g_1^{(2)}(k)|^2. \label{scasczakcor}
\end{eqnarray} 
Evidently, due to the constraint that $\gamma_{\text{SC},\text{inter}}$ can only takes  0 or $\pi$, it follows that as long as (\ref{scasczakcor}) is non-zero, $\gamma_{\text{LA},\text{inter}}$ must necessarily be fractional multiples of $\pi$ and non-quantized, as shown in Fig.~\ref{interzak}.
It is noteworthy that $\gamma_{\text{LA},\text{inter}}$ still exhibits discontinuities akin to $\gamma_{\text{SC},\text{inter}}$, both before and after the closure of the direct bandgap.
 
\section{Edge states} \label{edge}
The edge states, as localized at the boundaries, are special quantum states permitted within the bandgap and subject to topological protection. The BBC~\cite{Essin2011,Mong2011,Rhim2017,Rhim2018} suggests that topological invariants computed under PBC can predict number of edge states under OBC.
For example, in the SSH model, non-zero winding numbers/intercellular Zak phases correspond to the existence of a pair of edge states under OBC; 
if the winding numbers/intercellular Zak phase are zero, the system exhibits no edge states under OBC.
In this section, we separately check the edge states and BBC in the uneven ladder model for two configurations S.C. and L-As.C., respectively.
\subsection{IS-equipped S.C.} \label{sc}

\begin{figure}[h]
\includegraphics[width=7cm]{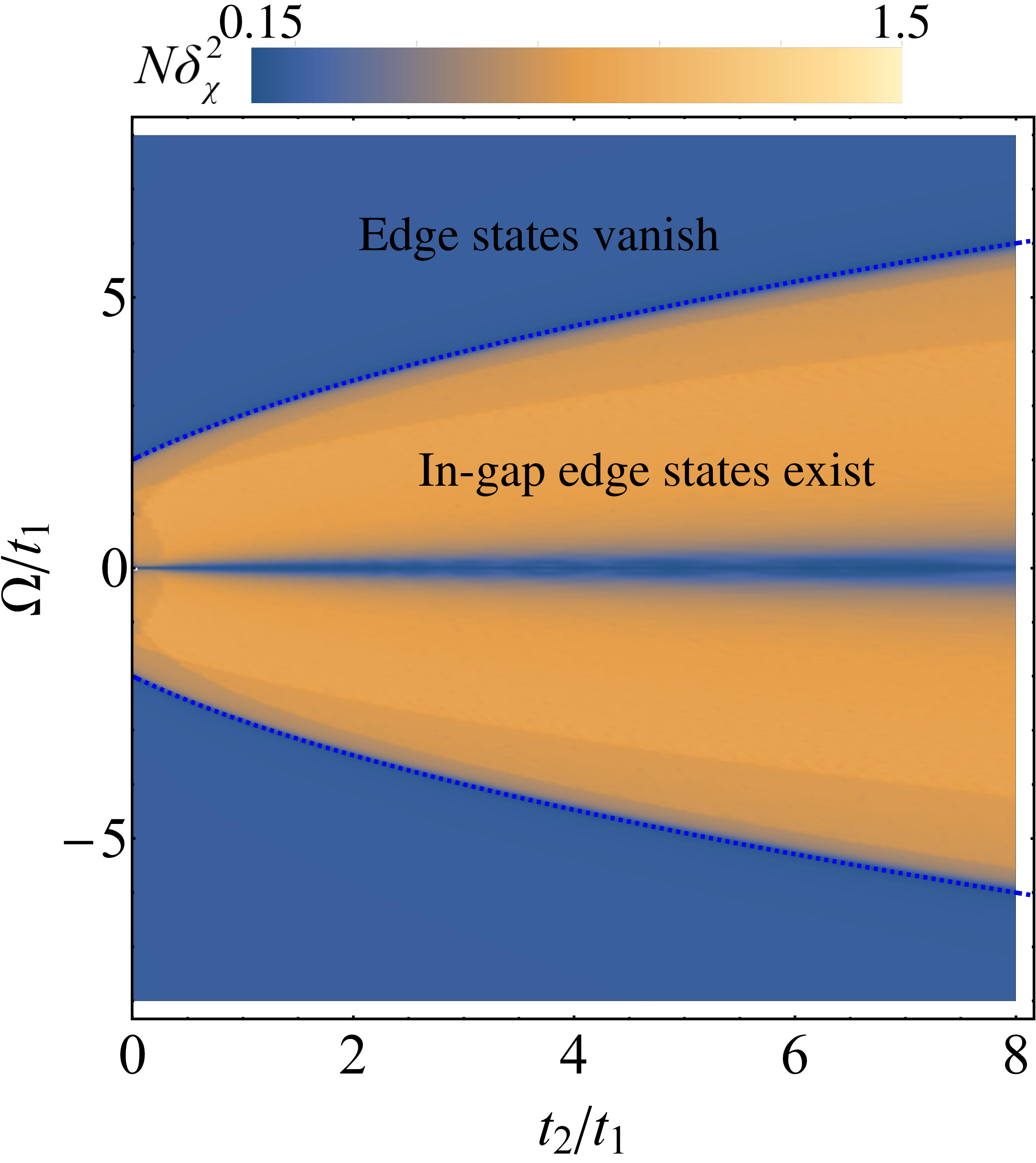}
\caption{BBC of S.C. for $t_2/t_1 >0$. $N\delta^2_{\chi}$ implies distribution of edge modes. 
The blue dashed line represents the closing of the direct bandgap. 
The region between the two blue dashed lines harbors edge states, while two bands have non-zero values of $\gamma_{\text{inter}}$ and $W_{\text{eff}}$. 
On either side of the two blue dashed lines, there are no edge states, while all three bands have $\gamma_{\text{inter}}$ and $W_{\text{eff}}$ equal to zero. 
Here, we have considered 90 unit cells. Due to finite size effects, the localization of edge states is not pronounced when $\Omega$ is relatively small (but non-zero), as indicated by the light blue shading near the abscissa.} \label{scfulledge}
\end{figure}

\begin{figure}[t]
\includegraphics[width=8.6cm]{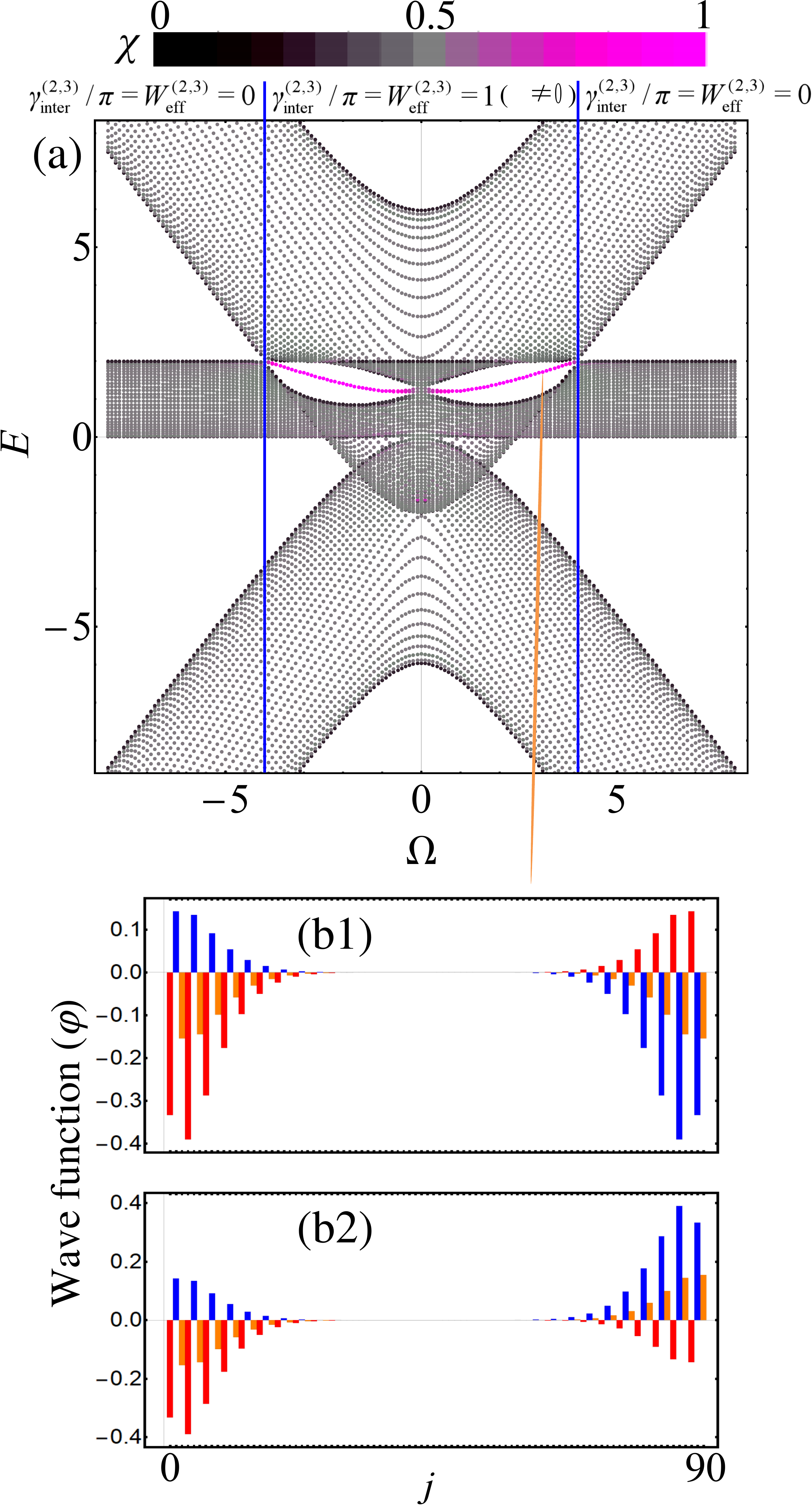}
\caption{OBC energy spectrum and edge states of S.C.. We set $t_1=1$, $t_2=3$, and adjust $\Omega$ to obtain the energy spectrum under OBC, as shown in (a). The colors in (a) indicate the $\chi^{i}$ of the $i$-th eigenstate to distinguish between bulk and edge states. The blue lines ($\Omega=\pm4$) mark the closing of the direct bandgap between the second and third bands. The reopening of this gap closure induces a sudden change in the intercellular Zak phase (effective winding number). (b1) and (b2) show a pair of degenerate edge states at $\Omega=3$ with $t_1=1$, $t_2=2$. In the bar plots, red, blue, and orange represent sublattices $A$, $B$, and $C$, respectively. Here, we use 30 unit cells.} \label{t1t2morethan0}
\end{figure}
To comprehensively investigate the edge modes, we define $\chi$ that can characterize the boundary-localized behavior of eigenstates $\varphi_i$ as
\begin{equation}
\chi^{(i)} = \sum_{j=1}^{N} \frac{\left|2j-N\right|}{N}\left | \varphi_i(j) \right | ^2,
\end{equation}
where $i$ is the index of eigenstate, $j$ denotes $j$-th site in real space.
If $\chi^{(i)}$ approach 1, it indicates the $i$-th eigenstate is an edge state. For a system of size $N$, we obtain a set $\{\chi^{(i)}\}$. We can calculate the variance $\delta^2_{\chi}$ of this dataset. Then,
\begin{equation}
N\delta^2_{\chi}=\sum_{i=1}^{N} (\chi^{(i)}-\bar{\chi})^2,
\end{equation}
where $\bar{\chi}$ is the average value of the set $\{\chi^{(i)}\}$. 
$N\delta^2_{\chi}$ can be used to determine whether the system exhibits boundary states: if the system is entirely extended, then $N\delta^2_{\chi}$ is small; if the system exhibits boundary states, then $N\delta^2_{\chi}$ is large.
We consider the case of $t_2/t_1 > 0$, noting that this is the condition satisfied by the uneven ladder model implemented in optical lattices. 

From Fig. \ref{scfulledge} and TABLE \ref{zakandwT}, we observed that, when $t_2/t_1>0$, the symmetry-protected S.C. follows the BBC: there are no edge states when intercelluar Zak phases (effective winding numbers) of the three bands are all zero; however, when there are non-zero intercelluar Zak phases (effective winding numbers), a pair of degenerate edge states emerge. 
In other words, when $\Omega<\left| 2 t_1 \sqrt{1+t_2/t_1}\right|$(obtained from the direct bandgap closing, i.e. (\ref{blueline})), edge states emerge.

Furthermore, from Fig. \ref{t1t2morethan0} we can see that the edge states are degenerate and in-gap, but their energy is non-zero. 
These non-zero edge modes are actually related to the lack of chiral symmetry in the system. 
A Hamiltonian with chiral symmetry, due to the property $CHC^\dagger=-H$, has a spectrum distribution with positive and negative symmetries (recalling the positive and negative symmetry of the spectrum in the SSH model protected by chiral symmetry). Therefore, the energies of the edge states within the bandgap of the two bands should also exhibit positive and negative symmetries, meaning $E^{\text{edge1}}_{\text{SSH}}=-E^{\text{edge2}}_{\text{SSH}}$. 
Thus, the energies of the two degenerate edge states should be zero, i.e., $E^{\text{edge1}}_{\text{SSH}}=E^{\text{edge2}}_{\text{SSH}}=0$. 
On the other hand, in the S.C., there is no chiral symmetry, and the two degenerate edge states are not required to satisfy the relationship of being opposite in sign, i.e., $E^{\text{edge1}}_{\text{SC}}\neq -E^{\text{edge2}}_{\text{SC}}$, thus naturally allowing the appearance of non-zero-energy edge states.

\subsection{IS broken L-As.C.\label{asc}} 

\begin{figure}[h]
\includegraphics[width=7cm]{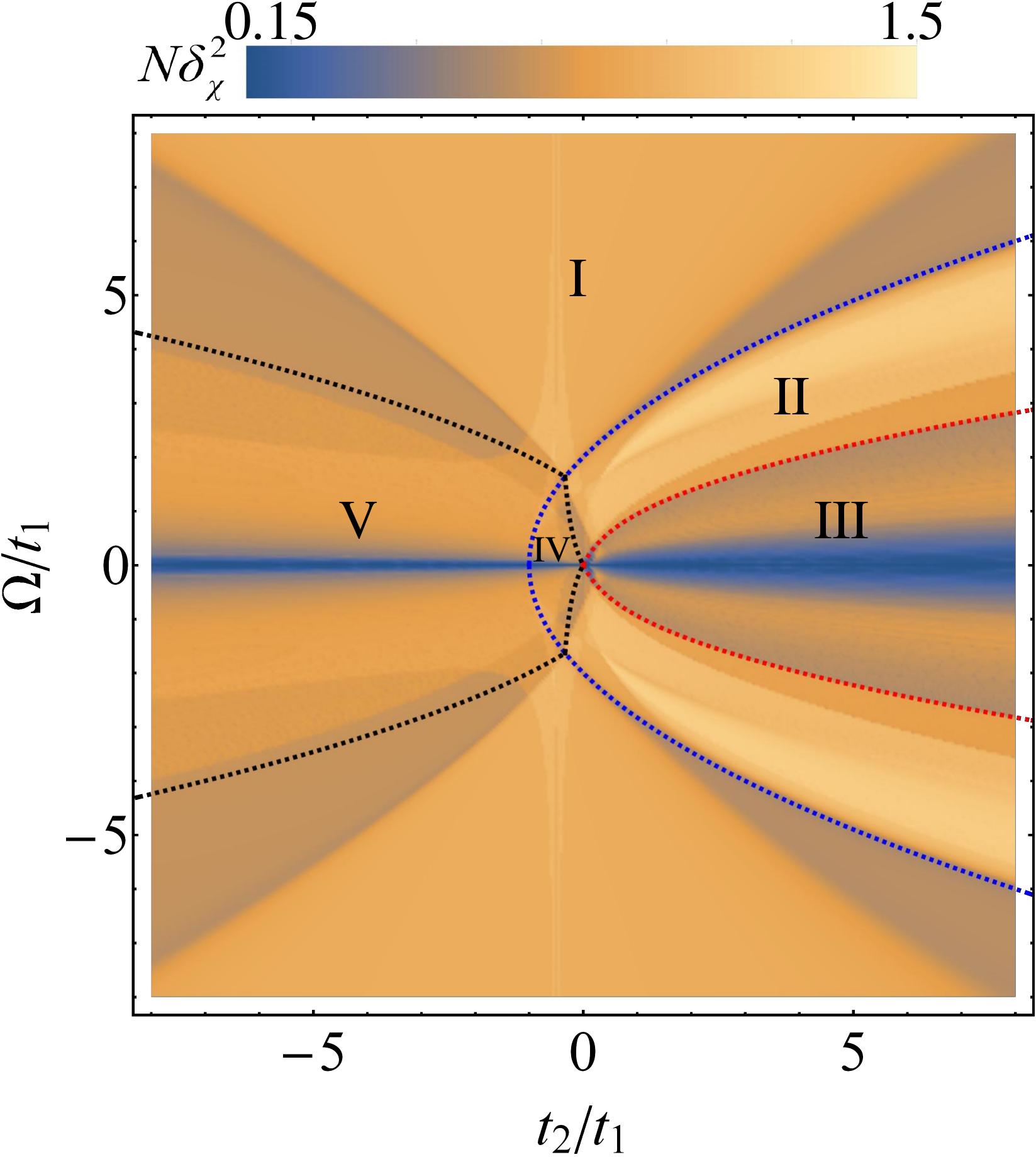}
\caption{The distribution of the edge modes of L-As.C.. The blue dashed lines denote the direct bandgap closure, while the red and black dashed lines mark distinct indirect bandgap closures. 
Here,  a system comprising 90 unit cells is employed. 
There exists relatively clear boundary states across the entire parameter space. 
For sufficiently small values of $\Omega$, due to finite size effects, boundary states are not prominent.
} \label{ascfulledge}
\end{figure}

L-As.C. shares the same bulk properties with S.C. (having identical band structures), differing only at the boundaries, where the distinct boundary configurations lead to S.C. possessing IS while L-As.C. does not. 
In this section, we demonstrate the impact of the broken IS at the boundaries on the edge states.

Aside from the scenario of weak coupling where $\Omega$ is extremely small (where the local characteristics of edge states are not prominent due to finite size effect), we observe that edge states of L-As.C. can persist consistently (as depicted in Fig. \ref{ascfulledge}). 
However, the quantity and local behavior of these edge states vary due to the influence of different bandgap closures (as evident from the magnitudes of $N\delta^2_\chi$ in Fig. \ref{ascfulledge} and the three dashed lines). Hereafter, we delve into the effects brought upon the edge states of L-As.C. by direct and indirect bandgap closures, respectively.

Due to the absence of IS and chiral symmetry protection, L-As.C. only possesses edge states localized at one-side of the boundary. 
We define average centroid position of their wavefunctions, defined as follows:

\begin{equation}
R^{(i)}=\sum_{j}^{N} \frac{j}{N} \left | \varphi_i(j) \right | ^2.
\end{equation}
Smaller $R$ indicates a more localized wave function at the left edge; 
conversely, larger $R$ signifies a more localized wave function at the right edge.

\begin{figure}[t]
\includegraphics[width=8.6cm]{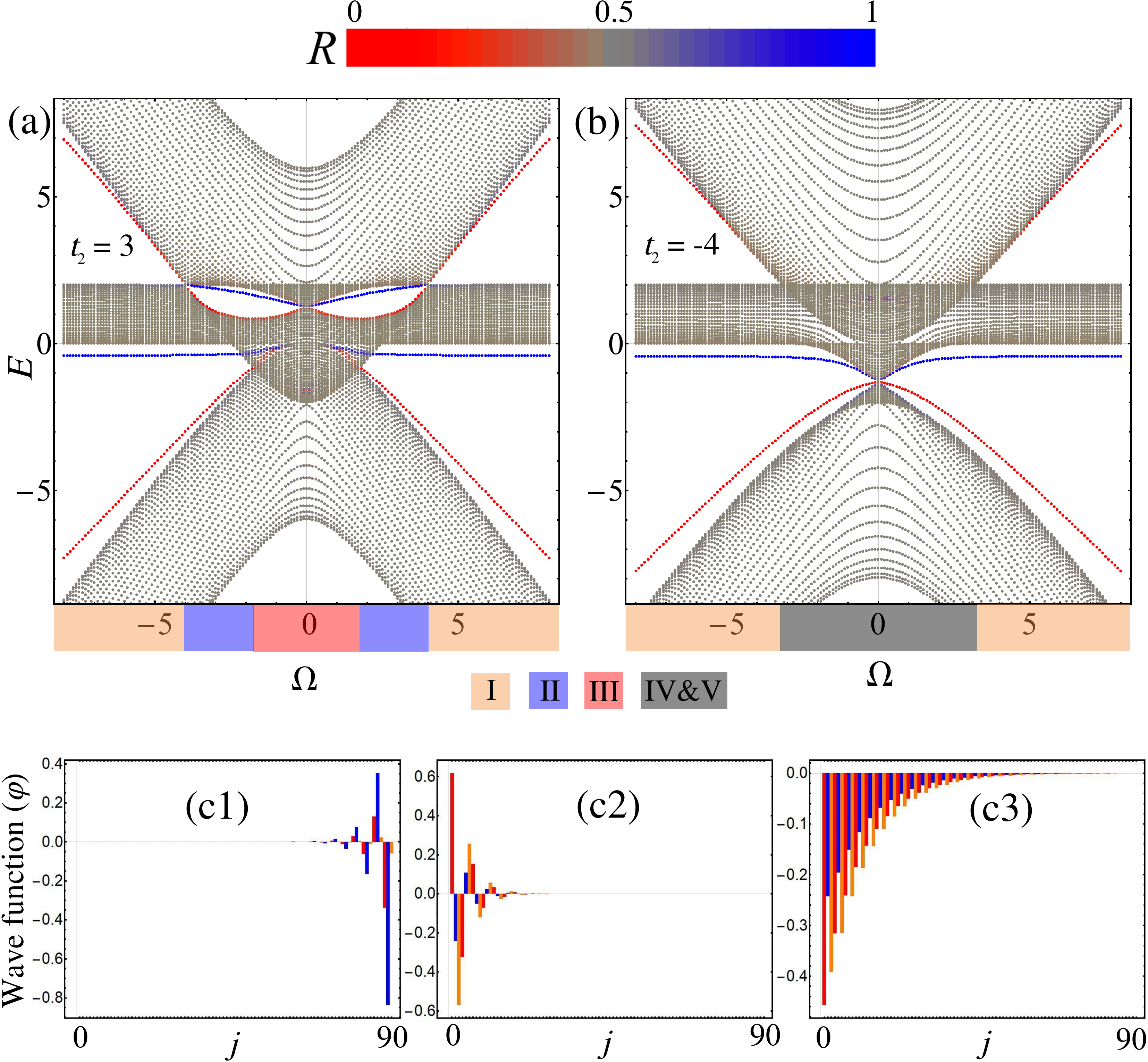}
\caption{(a-b) OBC energy spectrum  of L-As.C. for  $t_1=1$, $t_2=3$ (a), and $t_2=-4$ (b). 
$R$ approaching 0 (1) corresponds to edge states on the left (right) side.
(c1-c3) Edge state distribution within three bandgaps for $t_1=1$, $t_2=3$, and $\Omega=6$. 
Here, a system comprising 30 unit cells is employed.
} \label{ascfig}
\end{figure}

We first consider the case where $t_2/t_1 > 0$. The variation of the coupling strength $\Omega$ determines whether the system undergoes a closure and reopening of the direct bandgap and whether the indirect bandgap is closed. At any given point, the system may fall into one of the regions I, II, or III depicted in Fig. \ref{band structure}(a). We find distinct behaviors of edge states in these three regions (refer to Fig. \ref{ascfig}).

\begin{figure}[b]
\includegraphics[width=8.6cm]{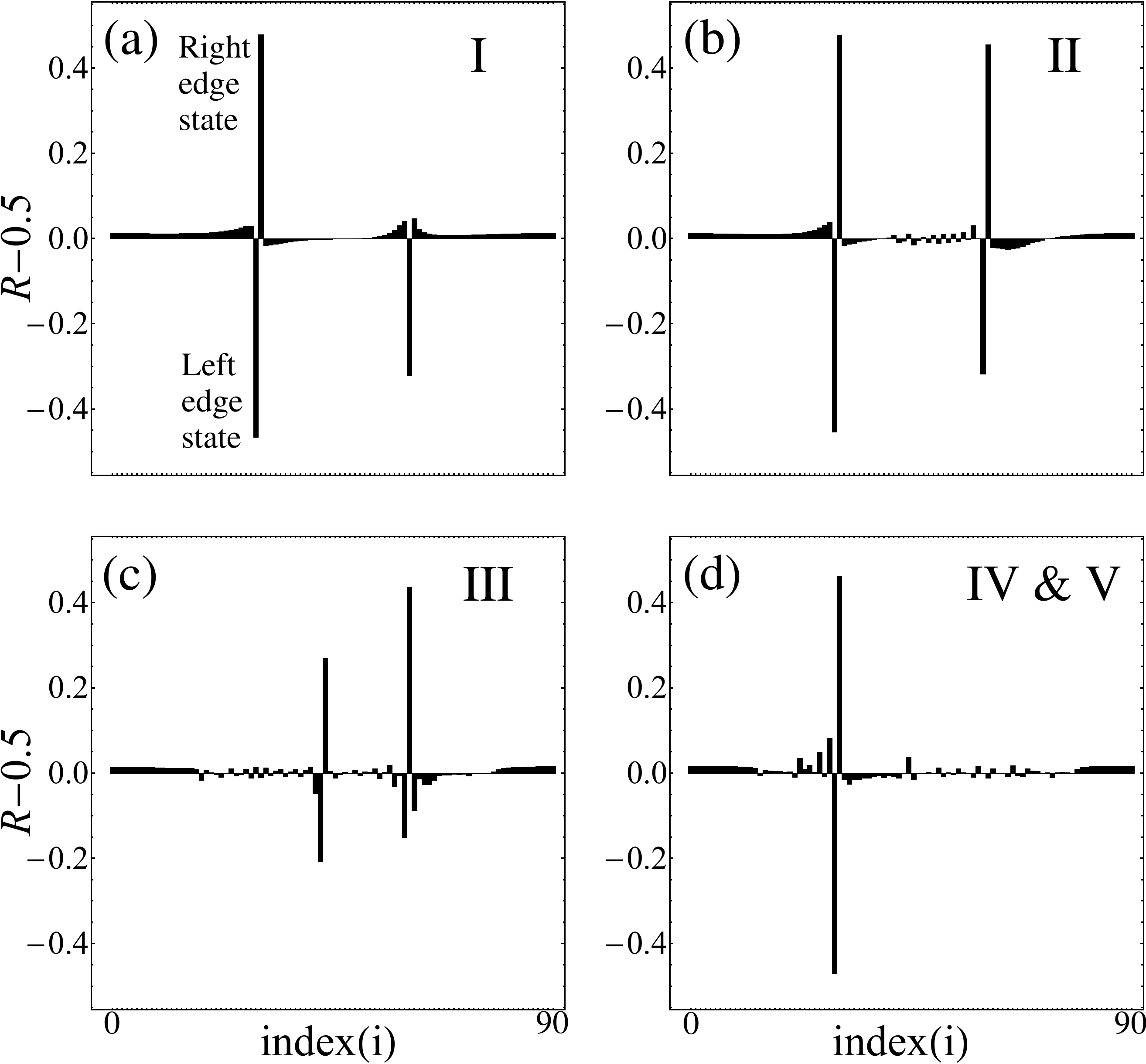}
\caption{The localized behavior of edge states of L-As.C. across different parameter regions. We take $t_2/t_1=3$, and $\Omega/t_1=5, 3, 1$ for (a), (b), (c), respectively, and take $t_2/t_1=-3$, $\Omega/t_1=1$ for (d), so that they represent the I, II, III, VI/V regions in Fig. \ref{band structure}. $R-0.5$ close to $-0.5$ $(0.5)$ means the local side is on the left (right). Here we use 30 unit cells, and let $t_1>0$.} \label{asedges}
\end{figure}

In region I, characterized by relatively large $\Omega$, the system hosts three in-gap edge states between the three bands. 
Here, the effective winding numbers $W^{(1,2,3)}_{\text{eff}}$ are $\{1, 1, 1\}$, as expected. When $\Omega$ is sufficiently large, all three bands of L-As.C. can be mapped to the non-trivial topological phase of the SSH model (Sec. \ref{projection}).

As $\Omega$ decreases gradually, the system undergoes a closure and reopening of the direct bandgap, falling into region II. 
Here, an additional right-sided localized edge state emerges (as observed in Fig. \ref{ascfig}(a) and compared in Fig. \ref{asedges}(a) and (b)). 
At this point, the effective winding numbers become $W^{(1,2,3)}_{\text{eff}} = \{0, 0, 1\}$.

Further reduction of $\Omega$ allows the system to enter region III, where, for $t_1 >0$, the system experiences an indirect bandgap closure between the first and second bands. 
This integration of the originally inter-bandgap edge states into bulk states weakens their boundary localization (akin to a critical mode blending between bulk and boundary states, as observed in Fig. \ref{ascfig}(a) and compared in Fig. \ref{asedges}(a) and (c)). 
Additionally, it is noteworthy that, despite the merging of edge states with bulk states, their energies remain continuous before and after merging.

We also consider the case of $t_2/t_1<0$. When $t_1>0$, the size of $\Omega$ determines whether the indirect bandgap between the second and third bands of the system closes. 
In region I, where this indirect bandgap is open, edge states between the three bands can coexist. 
In regions IV and V, where this indirect bandgap closes, the edge state originally present in the bandgap between the second and third bands will merge with bulk states, losing its edge-localized character (refer to Fig. \ref{ascfig}(b) and compare with Figs. \ref{asedges}(a) and (d)).

\section{extension of uneven ladders} \label{extension}
\begin{figure}[h]
\includegraphics[width=8.6cm]{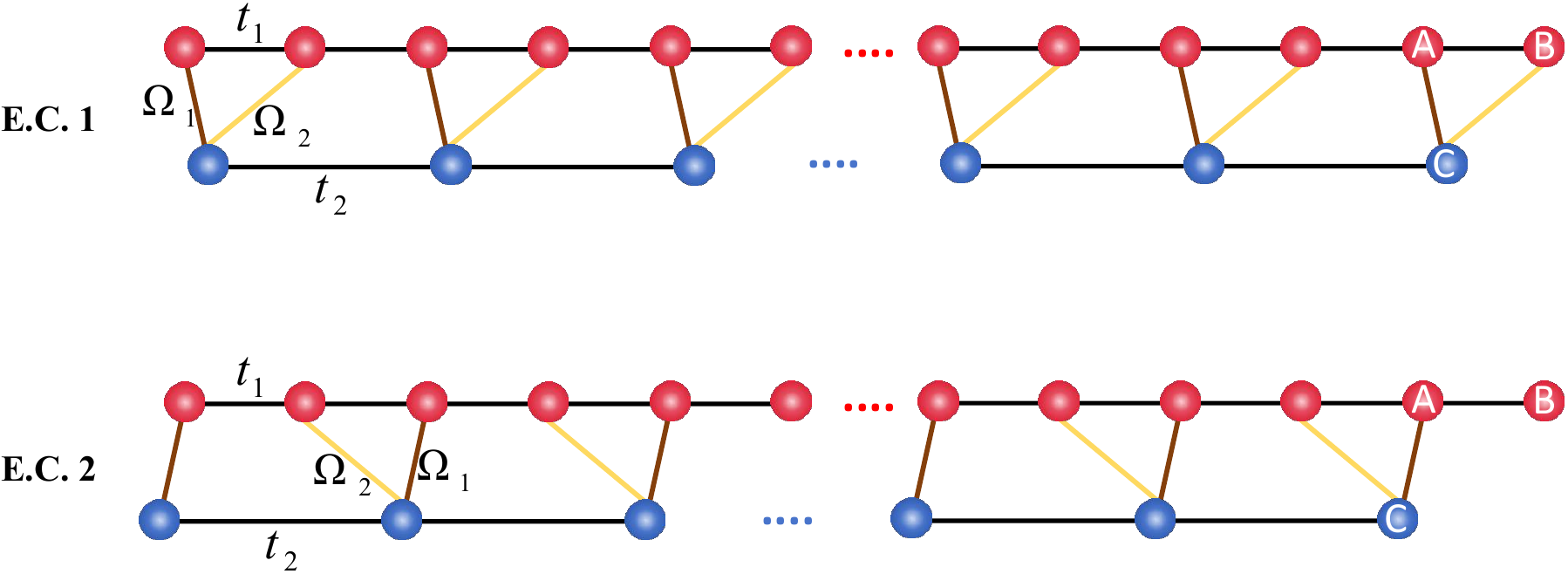}
\caption{Two configurations of uneven ladder model of extended configuration (E.C.).} \label{10}
\end{figure}

In this section, we further explore more general configurations and propose a simple implementation of a topological pump. 
The configurations we discussed so far have been rather coincidental, with the coupling strength of the $C$ sublattice hopping to the $A$ and $B$ sublattices being the same. 
To explore more general configurations, we extend the model as shown in Fig. \ref{10}. 
In these configurations, the coupling strengths of the two chains are denoted as $\Omega_1$ for nearest neighbor sites of sublattice $B$ and $C$, and $\Omega_2$ for nearest neighbor sites of sublattice $B$ and $C$, and $\Omega_1 \ne \Omega_2$. 
Consequently, IS will be broken. 

Specifically, we consider two extended configurations (referred as E.C. 1 and E.C. 2). E.C. 1 corresponds to a scenario based on S.C. where the lower lattice moves $(-a/4, 0)$, while E.C. 2 corresponds to a scenario based on S.C. where the lower lattice moves $(-a/2, -a/4)$, with $a$ being the lattice constant. 
It is noteworthy that we did not consider the lower chain's movement of $(0, a/2)$ as it is symmetrical with respect to $(-a/2, 0)$. 

The Bloch Hamiltonian of E.C. 1 reads 
\begin{equation}
   H^{\text{}}_{\text{E1}}(k) =\begin{pmatrix}
 0 & -t_1-t_1e^{-ik} & \Omega^{\text{E1}}_1  \\
-t_1-t_1e^{ik}  & 0 & \Omega^{\text{E1}}_2\\
 \Omega^{\text{E1}}_1 & \Omega^{\text{E1}}_2 & -2t_2 \cos k
\end{pmatrix}.
\end{equation}
Similarly, the Bloch Hamiltonian of E.C. 2 reads 
\begin{equation}
   H^{\text{}}_\text{E2}(k) =\begin{pmatrix}
 0 & -t_1-t_1e^{-ik} & \Omega^{\text{E2}}_1  \\
-t_1-t_1e^{ik}  & 0 & \Omega^{\text{E2}}_2 e^{ik}\\
 \Omega^{\text{E2}}_1 & \Omega^{\text{E2}}_2e^{-ik} & -2t_2 \cos k
\end{pmatrix}, 
\end{equation}
where we set lattice constant $a=1$.
Notice that if $\Omega^{\text{E1}}_1=\Omega^{\text{E2}}_1=\Omega_1$, $\Omega^{\text{E1}}_2=\Omega^{\text{E2}}_2=\Omega_2$ their band structures are identical due to $U H_{\text{E1}}(k) U^{\dagger}=H_{\text{E2}}(-k)$.

We can follow similar procedures as outlined in Sec. \ref{projection} to project this tripartite model onto the upper chain for E.C. 1, and the final result is
\begin{eqnarray}
\begin{aligned}
H_{\text{E1,eff}}^{(n)}(k)=&H_{AB}+\frac{1}{E_{\text{E1}}^{(n)}(k)-\delta }\begin{pmatrix}
 \Omega_1^2 &\Omega_1 \Omega_2 \\
 \Omega_1 \Omega_2 &\Omega_2^2 
\end{pmatrix}  \\
=&\frac{\frac{1}{2}(\Omega_1^2+\Omega_2^2)}{E_{\text{E1}}^{(n)}(k)+2t_2 \cos k }I  \\&+ \begin{pmatrix}
\Delta & -v-w e^{-ik} \\
 -v-we^{ik}&-\Delta
\end{pmatrix}, \label{e11}    
\end{aligned}
\end{eqnarray} where $E_{\text{E1}}^{(n)}(k)$ is the eigenvalue of $H^{\text{}}_{\text{E1}}(k)$ for $n$-th band, $v=t_1-\Omega_1\Omega_2/[E_{\text{E1}}^{(n)}(k)+2t_2\cos k)] $,  $w=t_1$, $\Delta=(\Omega_1^2-\Omega_2^2)/[E_{\text{E1}}^{(n)}(k)+2t_2\cos k)]$.
After ignoring the overall energy shift term, we obtain the effective projected Hamiltonian
\begin{eqnarray}
\tilde{H}_{\text{E1,eff}}^{(n)}(k)= \begin{pmatrix}
\Delta & -v-w e^{-ik} \\
 -v-we^{ik}&-\Delta
\end{pmatrix}, \label{e1}    
\end{eqnarray}
which can be regarded as a generalized RM model~\cite{Rice_Mele_1982}.

\begin{figure}[t]
\includegraphics[width=8.6cm]{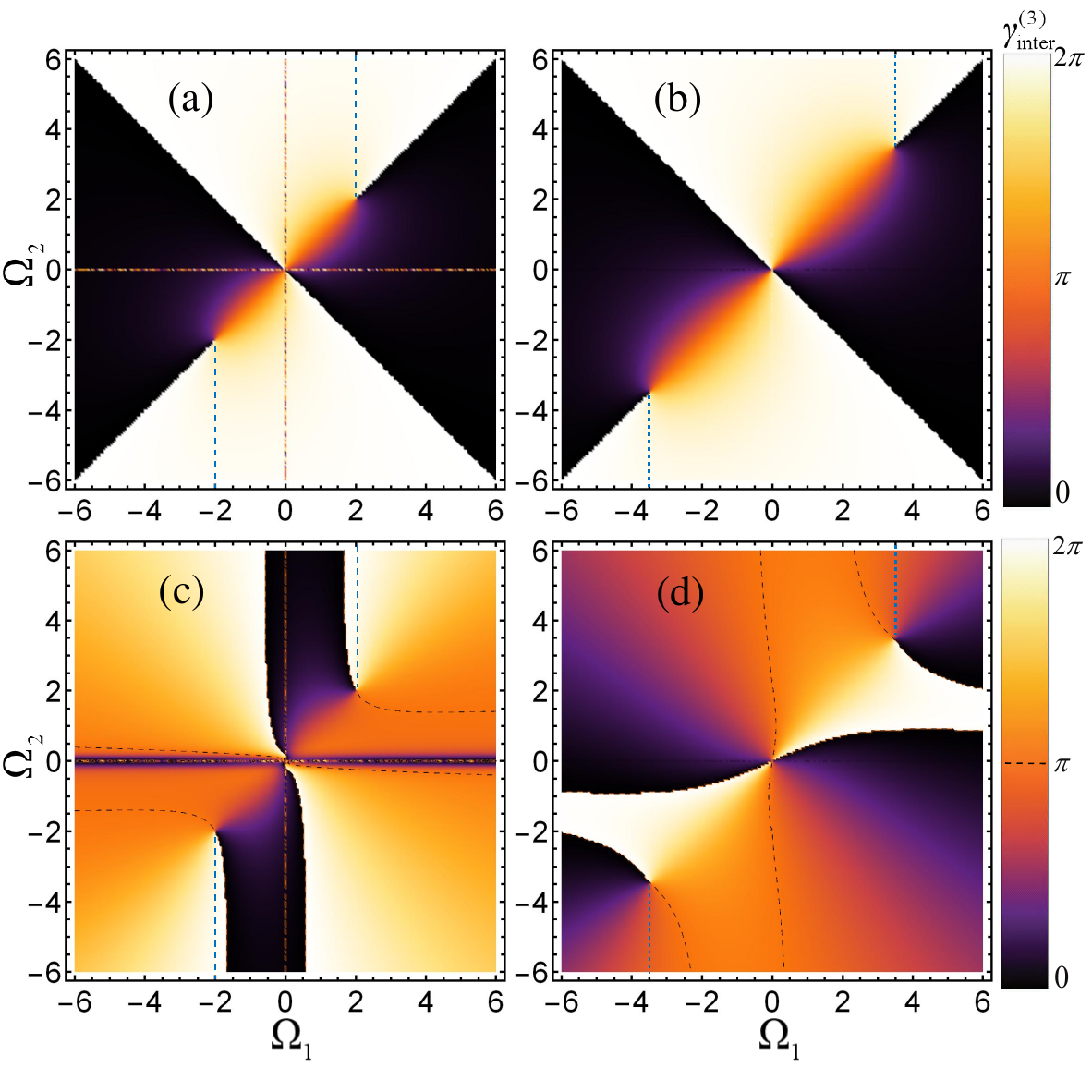}
\caption{(a-b) Intercelluar Zak phase of E.C. 1, $\gamma^{\text{E1},(3)}_{\text {inter }}$ v.s. $\Omega_1$, $\Omega_2$, for $t_2=0$, $2$, respectively. (c-d) Intercelluar Zak phase of E.C. 2, $\gamma^{\text{E2},(3)}_{\text {inter }}$ v.s. $\Omega_1$, $\Omega_2$, for $t_2=0$, $2$, respectively. The blue dashed lines in all of the four figures mark the direct gap closing points. The black dashed lines in (c), (d) mark value $\pi$.
Here $t_1=1$.
} \label{o1o2}
\end{figure}

Notice that due to the introduction of the $\sigma_z$ term in the effective projected Hamiltonian, the effective winding number cannot be defined; however, the intracellular Zak phase is always well-defined.
Subsequently, we calculate $\gamma^{(3)}_{\text {inter }}$ (mod $2\pi$) vs. $\Omega_1$, $\Omega_2$ for E.C. 1 and E.C. 2.
As shown in Fig. \ref{o1o2}(a) and (b) of E.C. 1, since $\Omega_1 \neq \Omega_2$ breaks the IS, the intracellular Zak phase can be a fractional multiple of $\pi$, whereas when $\Omega_1 = \Omega_2$, the system reduce to S.C. with 
integer multiple of $\pi$.
As shown in Fig. \ref{o1o2}(c) and (d) of E.C. 2, $\gamma^{\text{E2},(3)}_{\text {inter }}$ is a fractional multiple of $\pi$. 
However, as indicated by the black dashed lines, despite of the absence of symmetries protecting the topology, there are instances in E.C. 2 where the topological invariant $\gamma^{\text{E2},(3)}_{\text {inter }}=\pi$.

Furthermore, as shown in Fig. \ref{o1o2}, both $\gamma^{\text{E1},(3)}_{\text{inter}}$ and $\gamma^{\text{E2},(3)}_{\text{inter}}$ exhibit continuity near the direct gap closing point, with the value range spanning from 0 to $2\pi$. Therefore, by adjusting parameters to encircle the direct gap closing point, $\gamma^{(3)}_{\text{inter}}$ will inevitably undergo a change of $2\pi$. This brings to light an insight regarding topological pumps: if in some way we continuously alter the positional relationship between the upper and lower chains as well as other configurations, such that $\Omega_1$ and $\Omega_2$ vary continuously around the direct gap closing point, then such a closed path around the gap closing point will certainly result in an integer multiple change of $2\pi$ in the Zak phase. Since there is a correspondence between the Zak phase and the Wannier center, this implies that the system’s Wannier center will also shift by a quantized unit distance \cite{Citro_2016}.

\section{Conclusion}\label{conclusion}
We systematically investigate the topological phases and edge modes of the uneven ladder model. 
This model consists of two chains with different periods. We focus on the case where the period ratio is two. 
In the tight-binding limit, the uneven ladder is equivalent to a single chain with a three-site unit cell containing up to fourth-order hopping terms. 
We analyze its band structure and classify it based on whether the direct or indirect bandgaps are closed or open.

We propose a semi-analytical method to characterize its topological phases. 
This method projects the 1D ternary lattice system onto the subspace corresponding to the two sublattices, thereby constructing an effective two-component Bloch Hamiltonian. 
If two of the three sublattices exhibit exchange symmetry, the system can be projected onto a generalized SSH model, allowing the definition of effective winding numbers. 
For the uneven ladder model, the results obtained by this method coincide with numerical calculations of the Zak phase. Additionally, by breaking the symmetry of the sublattices, the three-component system can be projected onto a generalized RM model.

By selecting different unit cell arrangements, we construct two different configurations (S.C. and As.C.) and discuss their differences. 
These two configurations respectively preserve and break IS. While they are equivalent under PBC, they exhibit differences under OBC, particularly in the behavior of edge modes.
Configurations with IS can possess a pair of degenerate modes localized at the edges on both sides, corresponding to a topological phase transition where the direct band gap closes and reopens. 
On the other hand, configurations without IS have edge modes localized on one side only, which can persist before and after the closing-and-reopening of the direct band gap. 
Additionally, edge states can merge into bulk states due to the closure of the indirect band gap.

Building upon previous research, we decompose the Zak phase into intracellular and intercellular components, corresponding to surface bound charge and extra charge accumulation, respectively. We verify that both S.C. and As.C. configurations satisfy the aforementioned correspondence, and only the intercellular Zak phase of the S.C. configuration can serve as a $\mathbb{Z}_2$ topological invariant.

Furthermore, we investigate a more general uneven ladder model where the relative positions of the chains can vary, leading to richer topological properties. We find that the distribution of the Zak phase near the band closure points exhibits continuity, inspiring the construction of novel quantum topological pumps.

In summary, we have demonstrated the topological characteristics of the uneven ladder model, elucidating its potential as a novel system for exploring fundamental concepts in topological physics.

\begin{acknowledgments}
This work is supported by the Natural Science Foundation of Zhejiang Province, China (Grant Nos. LR22A040001, LY21A040004), and the National Natural Science Foundation of China (Grant No. 12074342, 11835011).
\end{acknowledgments}

\appendix
\section{Projective analysis\label{app_proj}}
 Consider a general Hermitian $3\times3$ Hamiltonian matrix written as $H=H_{AB}+H_{C}+H_{\Omega}$, where 
\begin{eqnarray}
H_{AB}&=&\begin{pmatrix}
 \alpha  & \Gamma  & 0\\
 \Gamma^* & \beta  & 0\\
  0& 0 &0
\end{pmatrix},\\
H_C&=&\begin{pmatrix}
  0&0  &0 \\
  0& 0 & 0\\
  0& 0 &\delta 
\end{pmatrix},\\ 
H_{\Omega} &=&\begin{pmatrix}
 0 & 0 &\Omega_1 \\
  0& 0 &\Omega_2 \\
 \Omega_1^* & \Omega_2^* &0
\end{pmatrix}.
\end{eqnarray}
These matrices are spanned in the space of $\left \{  \left | A  \right \rangle, \left | B  \right \rangle ,\left | C  \right \rangle \right \}$. 
Projection operators can be defined as
\begin{eqnarray}
\hat{P}_{AB}&=&\left | A  \right \rangle \left \langle A \right|  +\left | B  \right \rangle \left \langle B \right |,\\
\hat{P}_{C}&=&\left | C \right \rangle \left \langle C \right|,
\end{eqnarray}
which naturally satisfy $\hat{P}_{AB}+\hat{P}_{C}=I$, ${\hat{P}_{AB}^2}=\hat{P}_{AB}$, ${\hat{P}_{C}^2}=\hat{P}_{C}$.

We project the Hamiltonian into that of the sublattice with the Feshbach projection operator method which has widely been used in the few-body studies~\cite{Feshbach_1958,Feshbach_1962,Chin_2010}. We write down the eigenequation 
\begin{equation}
    H\left | \Psi_n  \right \rangle=E^{(n)}\left | \Psi_n  \right \rangle,
\end{equation}
where $n$ donates the $n$-th eigenstate. Then
\begin{eqnarray}
   \hat{P}_{AB}H(\hat{P}_{AB}+\hat{P}_{C}\hat{P}_{C}) \left | \Psi_n  \right \rangle&=&E^{(n)} \hat{P}_{AB}\left | \Psi_n  \right \rangle, \label{16}\\ 
\hat{P}_{C}H(\hat{P}_{AB}\hat{P}_{AB}+\hat{P}_{C}) \left | \Psi_n  \right \rangle&=&E^{(n)} \hat{P}_{C}\left | \Psi_n  \right \rangle.\label{proj}
\end{eqnarray}
From (\ref{proj}), we obtain 
\begin{equation}
  \hat{P}_{C} \left | \Psi_n  \right \rangle=(E^{(n)}-\hat{P}_{C}H\hat{P}_{C})^{-1}\hat{P}_{C}H\hat{P}_{AB}\hat{P}_{AB}\left | \Psi_n  \right \rangle.\label{18}
\end{equation}
Then we bring (\ref{18}) into (\ref{16}),
\begin{equation}
\begin{split}
     &(\hat{P}_{AB}H\hat{P}_{AB}-E^{(n)}\hat{P}_{AB}) \left | \Psi_n  \right \rangle=\\ 
     &-\hat{P}_{AB}H\hat{P}_{C}(E^{(n)}-\hat{P}_{C}H\hat{P}_{C})^{-1}\hat{P}_{C}H\hat{P}_{AB}\hat{P}_{AB}\left | \Psi_n  \right \rangle.
\end{split} \label{19}
\end{equation}
After deforming (\ref{19}), we  get 
\begin{equation}
   H_{\text{eff}}^{(n)}\hat{P}_{AB}\left | \Psi_n  \right \rangle =E^{(n)}\hat{P}_{AB}\left | \Psi_n  \right \rangle,
\end{equation}
where 
\begin{equation}
\begin{aligned}
H_{\text{eff}}^{(n)}=&\hat{P}_{AB}H\hat{P}_{AB}\\
    &+\hat{P}_{AB}H\hat{P}_{C}(E^{(n)}-\hat{P}_{C}H\hat{P}_{C})^{-1}\hat{P}_{C}H\hat{P}_{AB}.
\end{aligned}    
\end{equation}
Since $\hat{P}_{AB}H\hat{P}_{AB}=H_{AB}$, $\hat{P}_{C}H\hat{P}_{C}=\delta \left | C  \right \rangle \left \langle C \right |$, $\hat{P}_{AB}H_{\text{SC}}\hat{P}_{C}= (\Omega_1 \left | A  \right \rangle \left \langle C \right |+\Omega_2\left | B  \right \rangle \left \langle C \right |  )$, $\hat{P}_{C}H_{\text{SC}}\hat{P}_{AB}= (\Omega_1^*\left | C  \right \rangle \left \langle A \right |+\Omega_2^*\left | C  \right \rangle \left \langle B \right |)  $, we can rewrite ${H_{\text{eff}}^{(n)}}$ as
\begin{eqnarray}
\begin{aligned}
{H_{\text{eff}}^{(n)}}=&H_{AB}+G\begin{pmatrix}
 |\Omega_1|^2 &\Omega_1 \Omega_2^* \\
 \Omega_1^* \Omega_2 &|\Omega_2|^2 
\end{pmatrix}  \\
=&G\frac{|\Omega_1|^2+|\Omega_2|^2}{2}I  \\
&+ \begin{pmatrix}
\alpha+G\frac{|\Omega_1|^2-|\Omega_2|^2}{2} & \Gamma+G\Omega_1\Omega_2^*\\
 \Gamma^*+G\Omega_1^*\Omega_2&\beta-G\frac{|\Omega_1|^2-|\Omega_2|^2}{2}
\end{pmatrix},     
\end{aligned}
\end{eqnarray}
where $G=(E^{(n)}-\delta)^{-1}$.

For S.C. in Sec. \ref{projection}, $\alpha=\beta=0$, $\Gamma=-t_1-t_1 e^{-ik}$, $\delta=-t_2 \cos k$, $\Omega_1=\Omega_2=\Omega$, and $t_1$, $t_2$, $\Omega$ are real numbers, so that
\begin{eqnarray}
\begin{aligned}
H_{\text{SC,eff}}^{(n)}(k)=&H_{AB}+\frac{\Omega^2}{E_1^{(n)}(k)-\delta }\begin{pmatrix}
 1 &1 \\
 1 &1
\end{pmatrix}  \\
=&\frac{\Omega^2}{E_1^{(n)}(k)+2t_2 \cos k }I   \\
&+ \begin{pmatrix}
0 & -v-w e^{-ik} \\
 -v-we^{ik}&0
\end{pmatrix},    
\end{aligned}
\end{eqnarray}
where $v=t_1-\Omega^2E_1^{(n)}(k)+2t_2\cos k$,  $w=t_1$.

For E.C. 1 in Sec. \ref{extension}, $\alpha=\beta=0$, $\Gamma=-t_1-t_1 e^{-ik}$, $\delta=-t_2 \cos k$, $\Omega_1\ne\Omega_2$, and $t_1$, $t_2$, $\Omega_1$, $\Omega_2$ are real numbers, so that
\begin{eqnarray}
\begin{aligned}
H_{\text{E1,eff}}^{(n)}(k)=&H_{AB}+\frac{1}{E_{\text{E1}}^{(n)}(k)-\delta }\begin{pmatrix}
 \Omega_1^2 &\Omega_1 \Omega_2 \\
 \Omega_1 \Omega_2 &\Omega_2^2 
\end{pmatrix}  \\
=&\frac{\frac{1}{2}(\Omega_1^2+\Omega_2^2)}{E_{\text{E1}}^{(n)}(k)+2t_2 \cos k }I   \\
&+ \begin{pmatrix}
\Delta & -v-w e^{-ik} \\
 -v-we^{ik}&-\Delta
\end{pmatrix},    
\end{aligned}
\end{eqnarray}
where $v=t_1-\Omega_1\Omega_2/E_{\text{E1}}^{(n)}(k)+2t_2\cos k$,  $w=t_1$, $\Delta=(\Omega_1^2-\Omega_2^2)/[E_{\text{E1}}^{(n)}(k)+2t_2\cos k)]$.

\clearpage
\nocite{*}
\bibliographystyle{apsrev4-2}
\bibliography{ref}

\end{document}